\documentclass[a4paper,openright,8pt]{article}
\usepackage{fancyhdr}
\usepackage{graphicx}
\usepackage{amsfonts}
\usepackage{amsmath, amsthm, amssymb}
\usepackage[utf8]{inputenc}
\usepackage{dsfont}
\usepackage[english]{babel}
\usepackage{cite}
\usepackage[hyperindex=true,breaklinks=true,colorlinks=true,linkcolor=black,citecolor=blue]{hyperref}

\title{Multiphoton coherent states for bilayer graphene}
\author{David J. Fern\'andez C.\footnote{david.fernandez@cinvestav.mx} \,\ and Dennis I. Mart\'inez-Moreno\footnote{dmartinez@fis.cinvestav.mx}\\
  \small Physics Department, Cinvestav, P.O. Box 14-740, 07000 Mexico City, Mexico \\
  \date{September 1, 2022}
}

\begin{document}

\maketitle

\begin{abstract}
The multiphoton coherent states, a generalization to coherent sates, are derived for electrons in bilayer graphene placed in a constant homogeneous magnetic field which is orthogonal to the bilayer surface. For that purpose a generalized annihilation operator is constructed in order to determine the multiphoton coherent states as eigenstates of such operator with complex eigenvalue. In addition, some physical quantities are calculated for these states, as the Heisenberg uncertainty relation, probability density and mean energy value. Finally, in order to study the dynamics of the system the time evolution is explored and the time-correlation function is computed.        

\end{abstract}

\section{Introduction}
In 1900, Max Planck introduced for the first time the concept of quantization to explain black-body radiation. The revolutionary idea that the exchange of energy between radiation and matter takes place in a discrete way, through quantum units of energy, was the first breakthrough that gave rise to quantum mechanics, a probabilistic and indeterministic theory describing the microscopic world \cite{Z2009}. Since then, quantum mechanics has became the basis of modern physics and has been used in different branches, giving place to many theoretical and technological developments.
\\

Up to now, the efforts trying to establish a connection between quantum and classical theories keep constant, and they have contributed to the emergence of different semi-classical approaches. One of them is the so-called coherent states (CS) which were proposed first by Erwin Schr\"odinger in 1926 for the harmonic oscillator \cite{sc}. The standard coherent states (SCS) are quantum states that minimize the Heisenberg uncertainty relation, they evolve along the classical trajectory and are not deformed in time \cite{HR1987,G2009}. These are the reasons for the SCS to be sometimes called quasi-classical states, since they provide a natural framework to analyze the connection between quantum and classical mechanics.     
\\

One of the most famous applications of CS  happened in the early 1960s, when Glauber, Klauder and Sudarshan used them to describe coherent electromagnetic radiation \cite{glauber1,glauber2,glauber3,kl63a,kl63b,su63}, giving place to a new area in optics nowadays called quantum optics. The CS can be generalized, i.e., defined appropriately  in order to describe other systems in different areas of physics, as condensed matter, particle, nuclear and atomic physics, among other \cite{G2009,KS1985, AAG2014,Pe86}. In particular, the so-called multiphoton coherent states (MCS) \cite{BJQ1990,B1990,JB1993} are typically derived as eigenstates of powers of the annihilation operator; they have been addressed recently in \cite{MC19} for the harmonic oscillator in the framework of polynomial Heisenberg algebras\cite{fh99,fh02,cfnn04,BCF14,FM16}. Furthermore, the MCS were generated in \cite{DF19} for the supersymmetric harmonic oscillator.  
\\

It is worth noticing that coherent states and their generalizations have been employed recently to describe interesting physical systems that have attracted attention of the scientific community, the so-called 2D Dirac materials. \cite{DCR2019,DO2020}. In particular several works on the most conspicuous member of such a family, the graphene which is formed by carbon atoms arranged in a honeycomb hexagonal crystal lattice \cite{K2012}, have been addressed successfully through this semi-classical approach \cite{DF2017,DNN2019,DB20,CDO2020,MA2020,DNN21}. 
\\

Motivated by the pioneer work about coherent states for monolayer graphene \cite{DF2017}, a similar treatment was recently implemented to derive coherent states for electrons in bilayer graphene placed in a constant homogeneous magnetic field which is orthogonal to the bilayer surface \cite{FM20} (see also \cite{foc22}). With this in mind, the goal of this article is to extend and generalize the coherent states approach for bilayer graphene, by constructing now the corresponding multiphoton coherent states. For that purpose, this work has been organized as follows. In Sec. 2 a generalized annihilation operator is constructed and the MCS for bilayer graphene will be derived as eigenstates of such a matrix operator with complex eigenvalue $\tilde{\alpha}$. In order to describe and characterize the system, in Sec. 3 several physical quantities as the Heisenberg uncertainty relation, probability density and mean energy value will be determined. In Sec. 4 the time evolution is studied and the time-correlation function for the MCS will be obtained. Finally, in Sec. 5 the conclusions of this work are presented.

\section{Multiphoton coherent states}
As mentioned before, the MCS $\vert \tilde{z};m,j \rangle$ are generalizations of the standard CS which can be defined as eigenstates of a generalized or deformed annihilation operator $\hat{a}^{-}_{g}:= (\hat{a}^{-})^{m}$ with complex eigenvalues $\tilde{z}$,

\begin{equation}
\hat{a}^{-}_{g}\vert \tilde{z};m,j \rangle =\tilde{z}\vert \tilde{z};m,j \rangle, \,\ \qquad  \,\  \tilde{z}\in \mathbb{C}. \label{eq.mcs}
\end{equation}
By expressing $\vert \tilde{z},m,j \rangle$ as a superposition of Fock states, the MCS for the harmonic oscillator turn out to be

\begin{equation}
\vert \tilde{z};m,j \rangle= c_{j}^{m} \sum_{n = 0}^\infty \dfrac{\tilde{z}^{n}}{\sqrt{(mn+j)!}} \, \vert mn+j \rangle,\,\ \qquad  \,\  j=0,\ldots,m-1, \label{eq.mcs.osc}
\end{equation}
where $c_{j}^{m}$ are normalization constants. Note that these states are superposition of states $\vert mn+j \rangle$ with fixed $m$, $j$ and different $n$ whose difference of energy is an integer multiple of $m$, i.e., the number of photons required to jump between two levels of such superposition is always a multiple of $m$.

\subsection{MCS for bilayer graphene }
Bilayer graphene coherent states (BGCS) were built recently as eigenstates of the simplest diagonal annihilation operator $\hat{A}^{-}$ \cite{FM20}. Consider now a new generalized annihilation operator $\hat{A}^{-}_{g}$ defined as follows  

\begin{equation}
\hat{A}^{-}_{g}:=  (\hat{A}^{-})^{m}, \qquad m \in Z^{+}. \label{eq.operator}
\end{equation}
The MCS for bilayer graphene $\vert \tilde{\alpha};m \rangle$ can be constructed as eigenstates of $\hat{A}^{-}_{g}$  with complex eigenvalue $\tilde{\alpha}$,

\begin{equation}
\hat{A}^{-}_{g}\vert \tilde{\alpha};m \rangle =\tilde{\alpha}\vert \tilde{\alpha};m \rangle, \,\ \qquad  \,\  \tilde{\alpha}\in \mathbb{C}, \label{eq.mcsbg}
\end{equation}
where the states $\vert \tilde{\alpha};m \rangle$ are expressed as linear combinations of  $\left\lbrace \vert \Psi_{n} \rangle\right\rbrace ^{\infty}_{n=0}$, i.e.,

\begin{equation}
\vert \tilde{\alpha};m \rangle=\sum_{n = 0}^\infty C_{n}^{m} \, \vert \Psi_{n} \rangle, \label{eq.comb.lineal}
\end{equation}
with $\left\lbrace \vert \Psi_{n} \rangle\right\rbrace ^{\infty}_{n=0}$ being the eigenstates of the bilayer graphene effective Hamiltonian, whose explicit expressions turn out to be 

\begin{equation}
 \vert \Psi_{n} \rangle= \frac{\exp(iky)}{\sqrt{2^{1-\delta_{0n}-\delta_{1n}}}}
\left(\begin{array}{c}
 (1-\delta_{0n}-\delta_{1n}) \vert n-2 \rangle\\  \vert n \rangle \\
\end{array} 
\right), \,\ \,\ n=0,1,\ldots,  \label{eq.espinores} 
\end{equation}
where $\vert n \rangle$ are the harmonic oscillator Fock states.

\subsection{Generalized annihilation operator $\hat{A}^{-}_{g}$}

Let $\hat{A}^{-}_{g}$ the bilayer graphene generalized annihilator operator defined in Eq.(\ref{eq.operator}). Using the explicit expression of $\hat{A}^{-}$ given in \cite{FM20}, the operator $\hat{A}^{-}_{g}$ turns out to be 

\begin{equation}
\hat{A}^{-}_{g}=
\left(\begin{array}{cc}
f_{3}(\hat{N})f_{3}(\hat{N}+ \hat{1})\cdots f_{3}\left( \hat{N}+ (m-1)\hat{1}\right)  (\hat{a}^{-})^{m} & 0 \\ 0 & f(\hat{N} + \hat{1})f(\hat{N} + \hat{2}) \cdots f(\hat{N} + m\hat{1}) (\hat{a}^{-})^{m} \\
\end{array}
\right),  \label{eq.anni.operator}
\end{equation}
where $f$ and $f_{3}$ are two arbitrary functions of the number operator $\hat{N}$ which will be used to ensure that 

\begin{equation}
\hat{A}^{-}_{g}\vert \Psi_{n} \rangle=a_{n}\vert \Psi_{n-m} \rangle. \label{eq.eigen.anni.oper}
\end{equation}
By applying $\hat{A}^{-}_{g}$ on to the eigenstates $\vert \Psi_{n} \rangle$, the functions $f$ and $f_{3}$ must fulfil the following constraint,
\begin{equation}
\sqrt{(n-2)\cdots[n-(m+1)]}f_{3}(n-3) \cdots f_{3}\left( n-(m+2)\right) =\sqrt{n\cdots[n-(m-1)]} f(n)\cdots f\left( n-(m-1)\right), \label{eq.funciones}
\end{equation}
consequently, the generalized annihilation operator $\hat{A}^{-}_{g}$ can be rewritten as follows 
\begin{equation}
\hat{A}^{-}_{g}=
\left(\begin{array}{cc}
\frac{\sqrt{(\hat{N}+ \hat{3})\cdots\left( \hat{N}+ (m+2)\hat{1}\right) }}{\sqrt{(\hat{N} + \hat{1})\cdots(\hat{N} + m\hat{1})}}f(\hat{N}+\hat{3})\cdots f\left( \hat{N}+(m+2)\hat{1}\right)  (\hat{a}^{-})^{m} & 0 \\ 0 & f(\hat{N} + \hat{1})\cdots f(\hat{N} + m\hat{1})(\hat{a}^{-})^{m} \\
\end{array}
\right), \label{eq.anni.ope.expli}
\end{equation}
such that 
\begin{equation}
\hat{A}^{-}_{g}\vert \Psi_{n} \rangle=\left\{
	       \begin{array}{ll}
		 0 \,\ \, \, \, \qquad \qquad \qquad \qquad \qquad  \qquad \qquad \textrm {for}   & n=0, 1,\ldots, m-1, \\ \\
		 \sqrt{\dfrac{(1+\delta_{1m}) \, n!}{2}} \, [f(n)]! \, \vert \Psi_{0} \rangle \, \, \qquad \quad \,\ \,  \textrm{for} & n=m, \\ \\
		 \dfrac{\sqrt{n!}}{\sqrt{2}}\dfrac{[f(n)]!}{f(1)} \, \vert \Psi_{1} \rangle \,\ \, \qquad \quad  \qquad \qquad \,\ \, \textrm{for} & n=m+1, \\ \\
		 \sqrt{\dfrac{n!}{(n-m)!}}\dfrac{[f(n)]!}{[f(n-m)]!}\vert \Psi_{n-m} \rangle \qquad \textrm{for} & n=m+2, m+3,\ldots,
		 \end{array}
	     \right. \label{eq.accion}
\end{equation}
where $\delta_{ij}$ is the Kronecker delta and 

$$
[f(n)]!:=\left\{
	       \begin{array}{ll}
		 1 \,\  \,\ \,\ \,\ \,\ \,\ \,\ \,\ \,\ \,\ \,\ \,\  \textrm {for}   & n=0, \\
		 f(1) \cdot\cdot\cdot f(n) \,\ \,\ \textrm{for} & n>0.
		 \end{array}
	     \right.  
$$

\subsection{Bilayer graphene MCS as eigenstates of $\hat{A}^{-}_{g}$}

As mentioned before, the bilayer graphene MCS $\vert \tilde{\alpha};m \rangle$ can be constructed as eigenstates of the generalized annihilation operator $\hat{A}^{-}_{g}$ defined in Eq. (\ref{eq.anni.ope.expli}). Thus, from Eqs. (\ref{eq.mcsbg},\ref{eq.accion}) and using the linear independence of the states $\left\lbrace \vert \Psi_{n} \rangle\right\rbrace ^{\infty}_{n=0}$, two recurrence relationships for the coefficients $C_{n}^{m}$ are obtained, leading to 

\begin{equation}
C_{m}^{m}=\frac{\sqrt{2} \, \tilde{\alpha}}{\sqrt{(1+\delta_{1m}) \, m!}\,[f(m)]!}C_{0}^{m}, \label{eq.rec1}
\end{equation}
\begin{equation}
C_{n+m}^{m}=\sqrt{\dfrac{2^{\delta_{1n}}n!}{(n+m)!}}\frac{[f(n)]! \, \tilde{\alpha}^{n}}{[f(n+m)]!}C_{n}^{m}, \quad n=1,2,\ldots, \label{eq.rec2}
\end{equation}
Note that there are $m$ free parameters $\left\lbrace C_{0}^{m}, C_{1}^{m}, \ldots, C_{m-1}^{m} \right\rbrace=\left\lbrace C_{j}^{m}\right\rbrace_{j=0}^{m-1}$, thus $m$ independent sets of MCS $\vert \tilde{\alpha};m,j \rangle$ for bilayer graphene can be constructed, all of them depending on the particular choice of $f(n)$.\\

First of all, suppose that $f(n)\neq0 \quad \forall \quad n=1, 2,\ldots$ In particular, Eqs.(\ref{eq.rec1},\ref{eq.rec2}) for $m=1$ lead to

\begin{equation}
C_{n}^{1}=\sqrt{\dfrac{2^{1-\delta_{1n}}}{n!}}\frac{ \tilde{\alpha}^{n}}{[f(n)]!}C_{0}^{1}, \quad n=1,2,\ldots, \label{rec3}
\end{equation}
with $C_{0}^{1}$ being the only free parameter. Thus, the MCS for $m=1$ become

\begin{equation}
\vert \tilde{\alpha};1,0 \rangle=C_{0}^{1}\left[ \vert \Psi_{0} \rangle + \sum_{n = 1}^\infty \frac{\sqrt{2^{1-\delta_{1n}}} \, \tilde{\alpha}^{n} }{\sqrt{n!}\,[f(n)]!} \,  \vert \Psi_{n} \rangle\right], \label{eq.coherente1}
\end{equation} 
where $C_{0}^{1}$ will be used for normalizing them. Note that these states are identical to the BGCS derived in \cite{FM20} with $\tilde{\alpha}=\alpha$, i.e., for $m=1$ the BGCS are recovered.\\ 

For $m>1$ the $m$ independent relations resulting from Eqs. (\ref{eq.rec1},\ref{eq.rec2}) become

\begin{equation}
C_{mn+j}^{m}=\frac{\left[\sqrt{2}\delta_{0j}+\sqrt{(\delta_{1j}+j)!} \right] \, [f(j)]! \, \tilde{\alpha}^{n}}{\sqrt{(mn+j)!}\,[f(mn+j)]!}C_{j}^{m}, \quad n=1,2,\ldots, \label{eq.rec4}
\end{equation}
where $j=\left\lbrace 0, 1, 2, \ldots, m-1\right\rbrace$. From Eqs. (\ref{eq.comb.lineal}) and (\ref{eq.rec4}) the MCS turn out to be

\begin{equation}
\vert \tilde{\alpha};m, j \rangle=C_{j}^{m}\left[ \vert \Psi_{j} \rangle +  \sum_{n = 1}^\infty \frac{\left[\sqrt{2}\delta_{0j}+\sqrt{(\delta_{1j}+j)!} \right] \, [f(j)]! \, \tilde{\alpha}^{n}}{\sqrt{(mn+j)!}\,[f(mn+j)]!} \,  \vert \Psi_{mn+j} \rangle\right]. \label{eq.mcs}
\end{equation} 
The parameters $C_{j}^{m}$ will be used to normalize the states $\vert \tilde{\alpha};m, j \rangle$, which in general will depend on the values of the pair $\left\lbrace m, j \right\rbrace$. Some explicit expressions of MCS will be written next.

\paragraph{} For $m=2$ the index $j$ can take two values, $\left\lbrace 0, 1 \right\rbrace$, thus two sets of MCS will be obtained 
\begin{equation}
\vert \tilde{\alpha};2,0 \rangle=\left[1+2\sum_{n = 1}^\infty\dfrac{\vert \tilde{\alpha}\vert ^{2n}}{(2n)![[f(2n)]!]^{2}} \right]^{-1/2} \left[ \vert \Psi_{0} \rangle +  \sum_{n = 1}^\infty \frac{\sqrt{2}\tilde{\alpha}^{n} }{\sqrt{(2n)!}\,[f(2n)]!} \,  \vert \Psi_{2n} \rangle\right], \label{MULTI20}
\end{equation}

\begin{equation}
\vert \tilde{\alpha};2,1 \rangle=\left[1+2\sum_{n = 1}^\infty\dfrac{[f(1)]^{2} \vert \tilde{\alpha}\vert ^{2n}}{(2n+1)![[f(2n+1)]!]^{2}} \right]^{-1/2} \left[ \vert \Psi_{1} \rangle +  \sum_{n = 1}^\infty \frac{\sqrt{2}f(1)\tilde{\alpha}^{n}}{\sqrt{(2n+1)!}\,[f(2n+1)]!} \,  \vert \Psi_{2n+1} \rangle\right] . \label{MULTI21}
\end{equation}
These states can be obtained also as even and odd superpositions of bilayer graphene coherent states. This approach has been implemented recently for the particular case when $f(n)=1$ \cite{MA2021}.

\paragraph{}For $m=3$ the index $j$ can take three values, $\left\lbrace 0, 1, 2 \right\rbrace$, thus three sets of MCS are obtained

\begin{equation}
\vert \tilde{\alpha};3,0 \rangle=\left[1+2\sum_{n = 1}^\infty\dfrac{\vert \tilde{\alpha}\vert ^{2n}}{(3n)![[f(3n)]!]^{2}} \right]^{-1/2}\left[ \vert \Psi_{0} \rangle +  \sum_{n = 1}^\infty \frac{\sqrt{2}\tilde{\alpha}^{n} }{\sqrt{(3n)!}\,[f(3n)]!} \,  \vert \Psi_{3n} \rangle\right], \label{MULTI30}
\end{equation}

\begin{equation}
\vert \tilde{\alpha};3,1 \rangle=\left[1+2\sum_{n = 1}^\infty\dfrac{[f(1)]^{2} \vert \tilde{\alpha}\vert ^{2n}}{(3n+1)![[f(3n+1)]!]^{2}} \right]^{-1/2} \left[ \vert \Psi_{1} \rangle +  \sum_{n = 1}^\infty \frac{\sqrt{2}f(1)\tilde{\alpha}^{n}}{\sqrt{(3n+1)!}\,[f(3n+1)]!} \,  \vert \Psi_{3n+1} \rangle\right], \label{MULTI31}
\end{equation}

\begin{equation}
\vert \tilde{\alpha};3,2 \rangle=\left[1+2\sum_{n = 1}^\infty\dfrac{[f(1)f(2)]^{2} \vert \tilde{\alpha}\vert ^{2n}}{(3n+2)![[f(3n+2)]!]^{2}} \right]^{-1/2} \left[ \vert \Psi_{2} \rangle +  \sum_{n = 1}^\infty \frac{\sqrt{2}f(1)f(2)\tilde{\alpha}^{n}}{\sqrt{(3n+2)!}\,[f(3n+2)]!} \,  \vert \Psi_{3n+2} \rangle\right] . \label{MULTI32}
\end{equation}

Note that additional sets of MCS for bilayer graphene could be written explicitly, all of them depending on the particular choice of the function $f(n)$ and the parameters $\left\lbrace m, j \right\rbrace$. In order to analyze the electrons behavior in bilayer graphene, several physical quantities for the MCS will be computed in the following sections.   

\section{Physical quantities for the MCS}

The MCS $\vert \tilde{\alpha};m,j \rangle$ are quantum states belonging to a Hilbert space $\mathcal{H}$ from which several physical quantities can be extracted, in order to describe the system behavior in such approach.

\subsection{Heisenberg uncertainty relation}

One of the most important physical quantities useful to characterize a quantum state is the Heisenberg uncertainty relation (HUR). In order to obtain this quantity for the MCS, the following matrix operators $\hat{S}_{k}$ and their squares are defined as follows \cite{DCR2019}:

\begin{equation}
\hat{S}_{k}=\hat{s}_{k} \otimes \hat{1}, \qquad \hat{S}_{k}^{2}=\hat{s}_{k}^{2} \otimes \hat{1},
\end{equation} 
where
\begin{equation}
\hat{s}_{k}=\dfrac{1}{\sqrt{2}i^{k}}\left[\hat{a}^{-} + (-1)^{k}\hat{a}^{+} \right], 
\end{equation}
\begin{equation}
\hat{s}_{k}^{2}=\dfrac{1}{2}\left\lbrace 2\hat{N}+ \hat{1}+ (-1)^{k}\left[(\hat{a}^{-})^{2} + (\hat{a}^{+})^{2}\right]  \right\rbrace , 
\end{equation}
with $k=0, 1$, such that $\langle \hat{S}_{k}\rangle \vert_{k=0}=\langle \hat{q}\rangle$ and $\langle \hat{S}_{k}\rangle \vert_{k=1}=\langle \hat{p}\rangle$ (similarly for their squares). Thus, the mean values of these operators in the MCS (\ref{eq.mcs}) turn out to be:

\begin{equation}
\langle \hat{S}_{k}\rangle=0 \qquad \forall \,\ m,
\end{equation}

\begin{equation}
\begin{split}
\langle \hat{S}_{k}^{2}\rangle=&\vert C_{j}^{m}\vert^{2}\left\lbrace \dfrac{(1-\delta_{0j}-\delta_{1j})^{2}(2j-3) + 2j+1 }{2^{1-\delta_{0j}-\delta_{1j}}} + \left( \dfrac{[f(j)]! \, \left[\sqrt{2}\delta_{0j}+\sqrt{(\delta_{1j}+j)!} \right] }{[f(mn+j)]!}   \right)^{2} \right. \\ &
\left. \times \sum_{n = 1}^\infty\dfrac{\vert \tilde{\alpha}\vert^{2n} \, (2mn+2j-1)}{(mn+j)!}+(-1)^{k}2\Re(\tilde{\alpha})\left[\dfrac{[f(j)]!}{[f(mn+j)]! \, \sqrt{2^{1-\delta_{0j}-\delta_{1j}}\,j!}} + \right. \right. \\ &
\left. \left. + \left( \dfrac{[f(j)]! }{[f(mn+j)]!}   \right)^{2}  \left( \sum_{n = 1}^\infty\dfrac{\vert \tilde{\alpha}\vert^{2n}}{(mn+j)!}+ \sum_{n = 1}^\infty\dfrac{\vert \tilde{\alpha}\vert^{2n}}{\sqrt{(mn+j+2)! (mn+j-2)!}}   \right)    \right]\delta_{2m}   \right\rbrace .
\end{split}
\end{equation}

The standard deviation for $\hat{S}_{k}$ will be found through
\begin{equation}
\sigma_{\hat{S}_{k}}=\sqrt{\langle \hat{S}_{k}^{2}\rangle - \langle \hat{S}_{k}\rangle^{2}}, 
\end{equation}
thus the HUR can be obtained for the MCS, which is given by

\begin{equation}
\sigma_{\hat{q}_{\tilde{\alpha}}}\sigma_{\hat{p}_{\tilde{\alpha}}}\geq\dfrac{1}{2}. \label{HUR}
\end{equation}

\begin{figure}[!]
\begin{center} 
  \includegraphics[width=10cm]{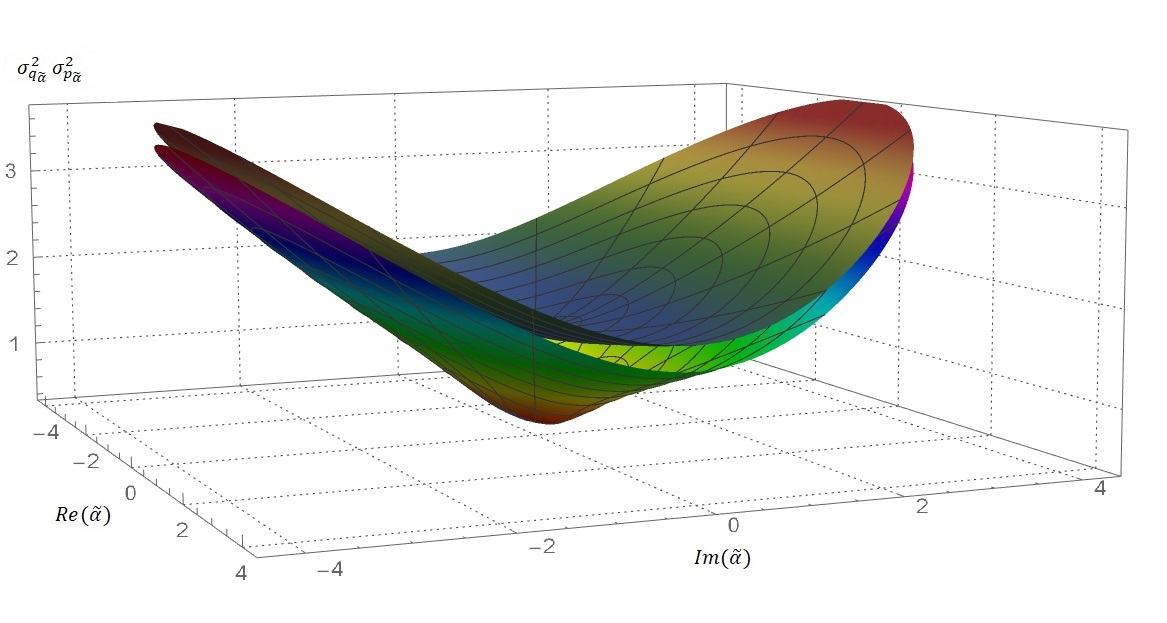}
   \caption{Heisenberg uncertainty product $\sigma_{\hat{q}_{\tilde{\alpha}}}\sigma_{\hat{p}_{\tilde{\alpha}}}$ as function of $\tilde{\alpha}$ for the MCS with $m=2$, $j=0,1$ and $f(n)=1$. }
     \label{imagen-Heisenberg2}
  \end{center}  
\end{figure}

\begin{figure}[!]
\begin{center} 
  \includegraphics[width=10cm]{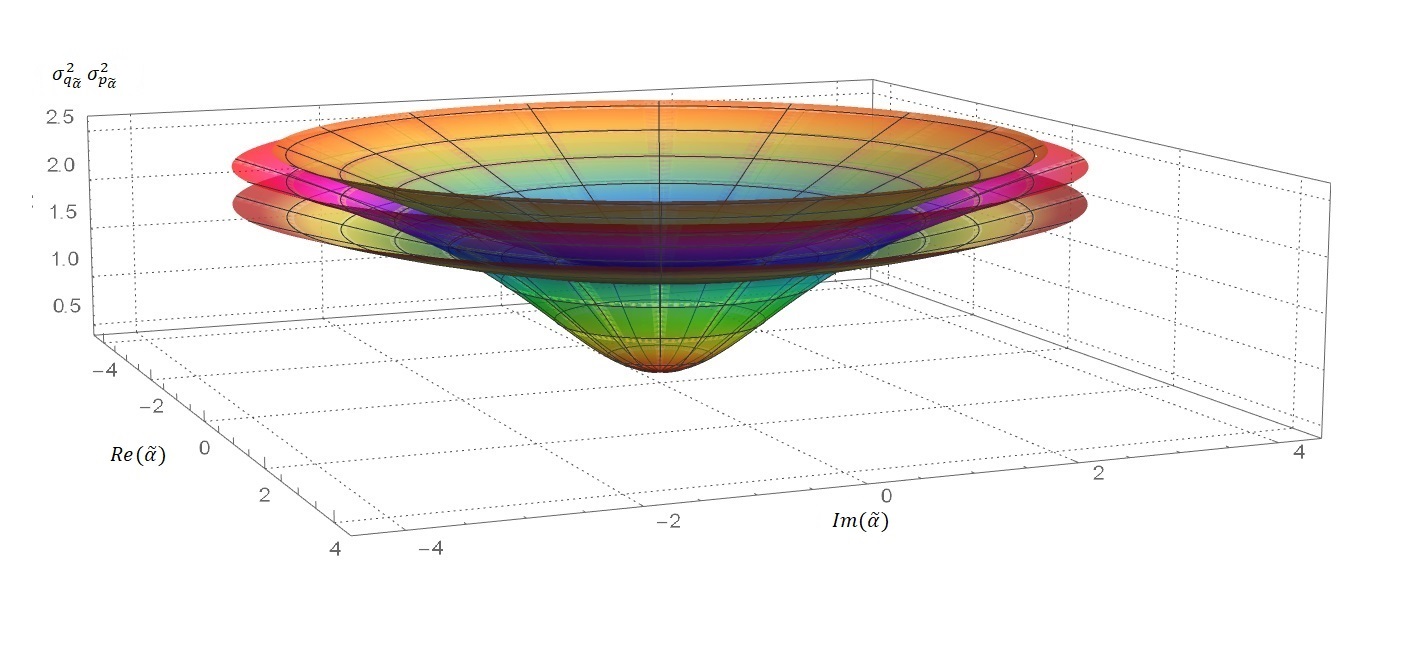}
   \caption{Heisenberg uncertainty product $\sigma_{\hat{q}_{\tilde{\alpha}}}\sigma_{\hat{p}_{\tilde{\alpha}}}$ as function of $\tilde{\alpha}$ for the MCS with $m=3$, $j=0,1,2$ and $f(n)=1$. }
     \label{imagen-Heisenberg3}
  \end{center}  
\end{figure}

Figures \ref{imagen-Heisenberg2} and \ref{imagen-Heisenberg3} show the resulting Heisenberg uncertainty product $\sigma_{\hat{q}_{\tilde{\alpha}}}\sigma_{\hat{p}_{\tilde{\alpha}}}$ for the MCS as function of $\tilde{\alpha}$ with $f(n)=1$ and the two values $m=\left\lbrace 2,3 \right\rbrace$ respectively.

\subsection{Probability density}
The MCS probability density is determined as

\begin{equation}
\rho_{\tilde{\alpha}}(x)= \Psi_{\tilde{\alpha}}^{\dagger}(x,y) \, \Psi_{\tilde{\alpha}}(x,y), 
\end{equation}
where $\Psi_{\tilde{\alpha}}(x,y)=\langle x,y \vert \tilde{\alpha};m,j\rangle$.
\\

This quantity represents the position probability density, i.e., $\rho_{\tilde{\alpha}}(x)\mathrm{d}x$\footnote{Although the wave function $\Psi_{\tilde{\alpha}}(x,y)$ depends explicitly on $x$ and $y$, the associated probability density is independent of $y$ due to the translational symmetry along this direction; in addition, it will be time independent for stationary states.} is the probability of finding the electron between $x$ and $x + \mathrm{d}x$. For the MCS of Eq. (\ref{eq.mcs}) the probability density becomes 

%%%%%%%%%%%%%%%%%%%%%%%%%%%%%%%%%%%%%%%%%%%%%%%%%%%%%%%%%%%%%%%%%%%%%%%%%%%%%%%%%%%%%%%%%%%%%%%%%%%%%%%%%%%%%%%%%%%%%%%%%%%%%%%%%%%%%%%%%%%%%%%%%%

\begin{figure}[!]
\begin{center} 
  \includegraphics[width=16.5cm]{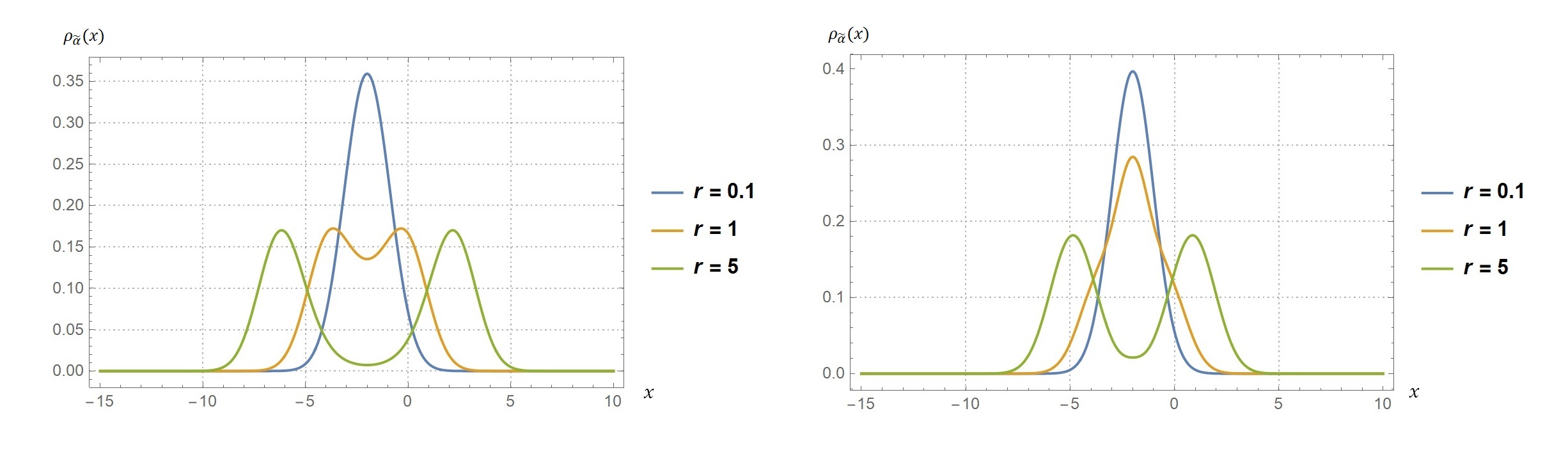}
   \caption{Probability density $\rho_{\tilde{\alpha}}(x)$ for the MCS with $f(n)=1$, $\omega_{c}^{*}=1$, $m=2$ and $j=0$. Different values of $r=\vert \tilde{\alpha} \vert$ are taken for $\theta=0$ (left) and $\theta= \pi/2$ (right). }
     \label{imagen-density20}
  \end{center}  
\end{figure} 

%%%%%%%%%%%%%%%%%%%%%%%%%%%%%%%%%%%%%%%%%%%%%%%%%%%%%%%%%%%%%%%%%%%%%%%%%%%%%%%%%%%%%%%%%%%%%%%%%%%%%%%%%%%%%%%%%%%%%%%%%%%%%%%%%%%%%%%%%%%%%%%%%%

\begin{equation}
\begin{split}
 &\rho_{\tilde{\alpha}}(x)=\vert C_{j}^{m}\vert^{2} \left\lbrace \dfrac{(1-\delta_{0j}-\delta_{1j})^{2}\vert \psi_{j-2}\vert^{2}+\vert \psi_{j}\vert^{2}}{\sqrt{2^{1-\delta_{0j}-\delta_{1j}}}} + \displaystyle \dfrac{{[\sqrt{2}\delta_{0j}+\sqrt{(\delta_{1j}+j)!}]}}{\sqrt{2^{2-\delta_{0j}-\delta_{1j}}}} 2\Re \left[\sum_{n=1}^\infty \frac{\tilde{\alpha}^{n}}{\sqrt{(mn+j)!}}   \right. \right. \\ &
 \left.\left. \times \frac{[f(j)]!}{[f(mn+j)]!} \left( (1-\delta_{0j}-\delta_{1j})\psi_{mn+j-2} \psi_{j-2}^{*} + \psi_{mn+j}\psi_{j}^{*} \right) \right] + \dfrac{[\sqrt{2}\delta_{0j}+\sqrt{(\delta_{1j}+j)!}]^{2}}{2} \left | \sum_{n=1}^\infty \frac{\tilde{\alpha}^{n}}{\sqrt{(mn+j)!}} \right. \right. \\&
\left. \left. \times \frac{[f(j)]!}{[f(mn+j)]!} \, \psi_{mn+j-2} \right |^{2}+ \dfrac{[\sqrt{2}\delta_{0j}+\sqrt{(\delta_{1j}+j)!}]^{2}}{2} \left | \sum_{n=1}^\infty \frac{\tilde{\alpha}^{n}}{\sqrt{(nm+j)!}} \frac{[f(j)]!}{[f(mn+j)]!} \, \psi_{mn+j} \right |^{2}   \right\rbrace. \label{eq.prob1}
\end{split}
\end{equation}
Using the polar form $\tilde{\alpha}=re^{i\theta}=r\left( \cos\theta+i\sin\theta\right)$, the previous probability density turns out to be

\begin{equation}
\begin{split}
 &\rho_{\tilde{\alpha}}(x)=\vert C_{j}^{m}\vert^{2} \left\lbrace \dfrac{(1-\delta_{0j}-\delta_{1j})^{2}\vert \psi_{j-2}\vert^{2}+\vert \psi_{j}\vert^{2}}{\sqrt{2^{1-\delta_{0j}-\delta_{1j}}}} +  \dfrac{{[\sqrt{2}\delta_{0j}+\sqrt{(\delta_{1j}+j)!}]}}{\sqrt{2^{2-\delta_{0j}-\delta_{1j}}}} \left[\sum_{n=1}^\infty \frac{r^{n}\cos(n\theta)}{\sqrt{(mn+j)!}}  \right. \right. \\ &
 \left.\left. \times \frac{[f(j)]!}{[f(mn+j)]!} \left( (1-\delta_{0j}-\delta_{1j})\psi_{mn+j-2} \psi_{j-2}^{*} + \psi_{mn+j}\psi_{j}^{*}\right) + \sum_{k=1}^\infty \frac{r^{k}\cos(k\theta)}{\sqrt{(mk+j)!}} \,\frac{[f(j)]!}{[f(mk+j)]!}  \right. \right. \\&
\left. \left. \times \left( (1-\delta_{0j}-\delta_{1j}) \psi_{mk+j-2}^{*} \psi_{j-2} + \psi_{mk+j}^{*} \psi_{j}\right)\right] + \dfrac{[\sqrt{2}\delta_{0j}+\sqrt{(\delta_{1j}+j)!}]^{2}}{2}\sum_{n=1}^\infty\sum_{k=1}^\infty \frac{r^{n+k} \cos[(n-k)\theta]}{\sqrt{(mn+j)!(mk+j)!}}  \right. \\&
\left. \times \frac{[[f(j)]!]^{2}(\psi_{mn+j-2} \psi_{mk+j-2}^{*} + \psi_{mn+j}\psi_{mj+j}^{*} )}{[f(mn+j)]![f(mk+j)]!} \right\rbrace.  \label{eq.prob11}
\end{split}
\end{equation}
 
\begin{figure}[!]
\begin{center} 
  \includegraphics[width=16.5cm]{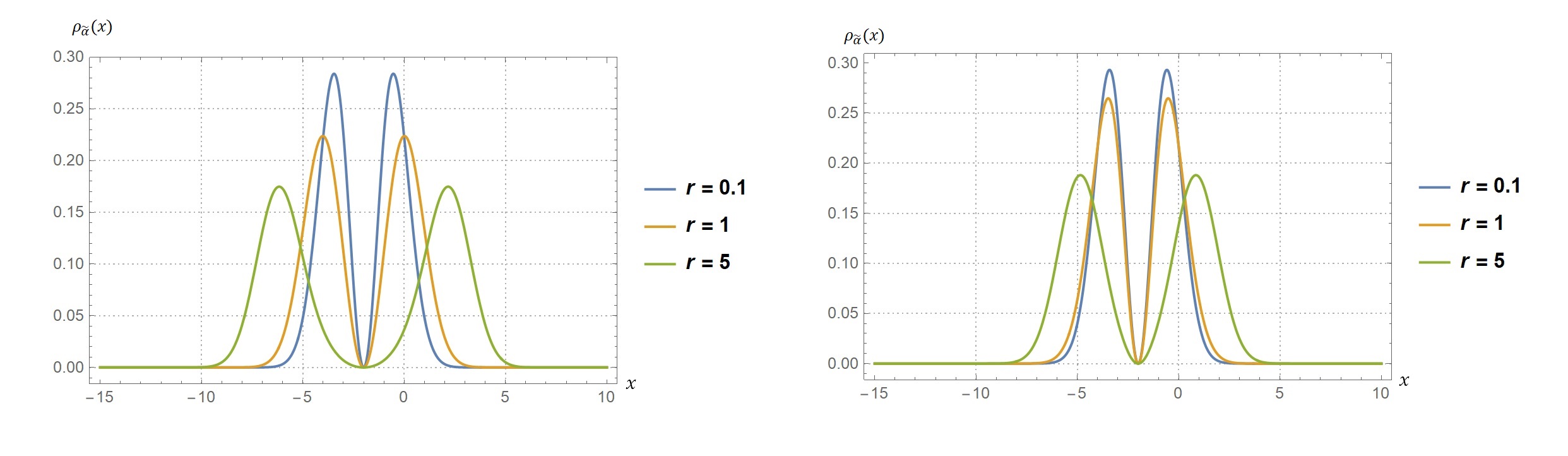}
   \caption{Probability density $\rho_{\tilde{\alpha}}(x)$ for the MCS with $f(n)=1$, $\omega_{c}^{*}=1$, $m=2$ and $j=1$. Different values of $r=\vert \tilde{\alpha} \vert$ are taken for $\theta=0$ (left) and $\theta= \pi/2$ (right). }
     \label{imagen-density21}
  \end{center}  
\end{figure}

\begin{figure}[!]
\begin{center} 
  \includegraphics[width=16.5cm]{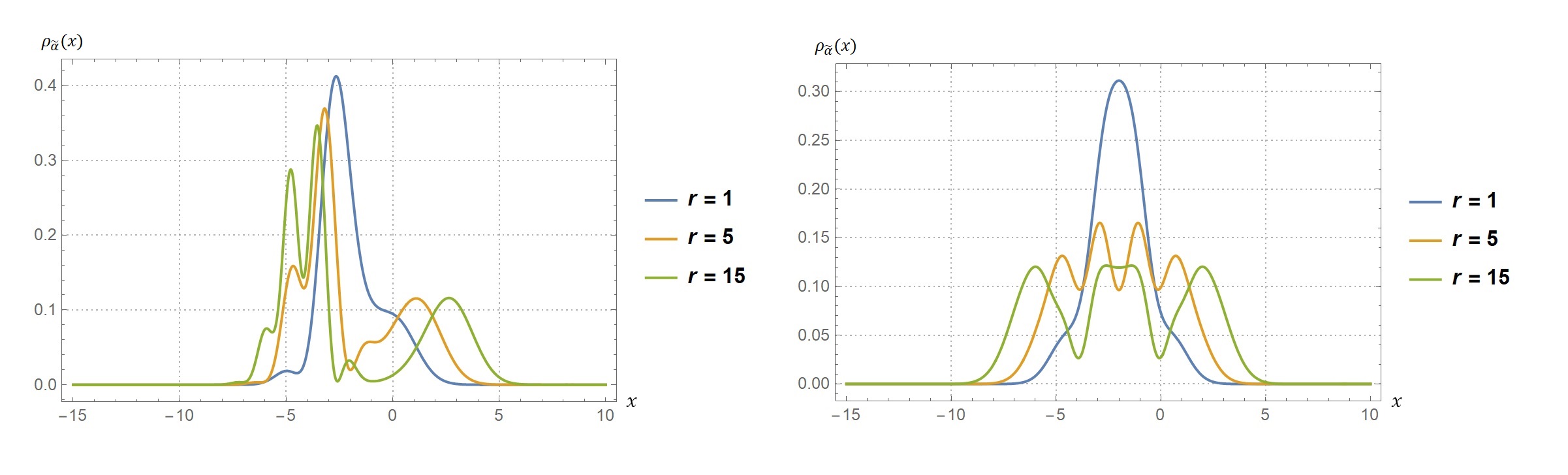}
   \caption{Probability density $\rho_{\tilde{\alpha}}(x)$ for the MCS with $f(n)=1$, $\omega_{c}^{*}=1$, $m=3$ and $j=0$. Different values of $r=\vert \tilde{\alpha} \vert$ are taken for $\theta=0$ (left) and $\theta= \pi/2$ (right). }
     \label{imagen-density30}
  \end{center}  
\end{figure}

\begin{figure}[!]
\begin{center} 
  \includegraphics[width=16.5cm]{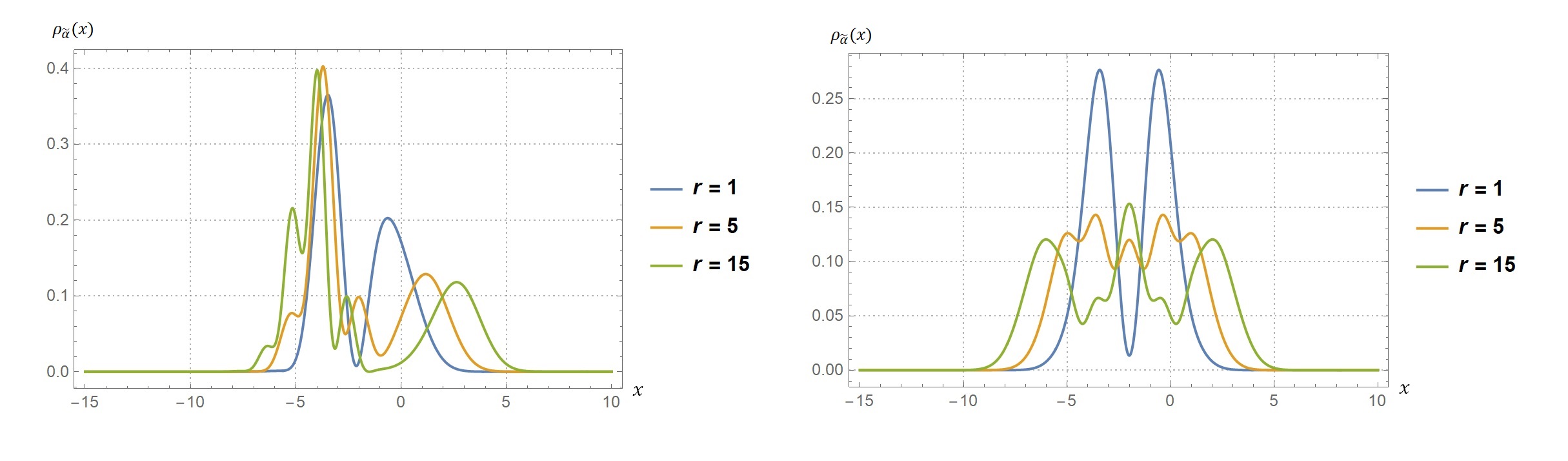}
   \caption{Probability density $\rho_{\tilde{\alpha}}(x)$ for the MCS with $f(n)=1$, $\omega_{c}^{*}=1$, $m=3$ are $j=1$. Different values of $r=\vert \tilde{\alpha} \vert$ are taken for $\theta=0$ (left) and $\theta= \pi/2$ (right). }
     \label{imagen-density31}
  \end{center}  
\end{figure}

\begin{figure}[!]
\begin{center} 
  \includegraphics[width=16.5cm]{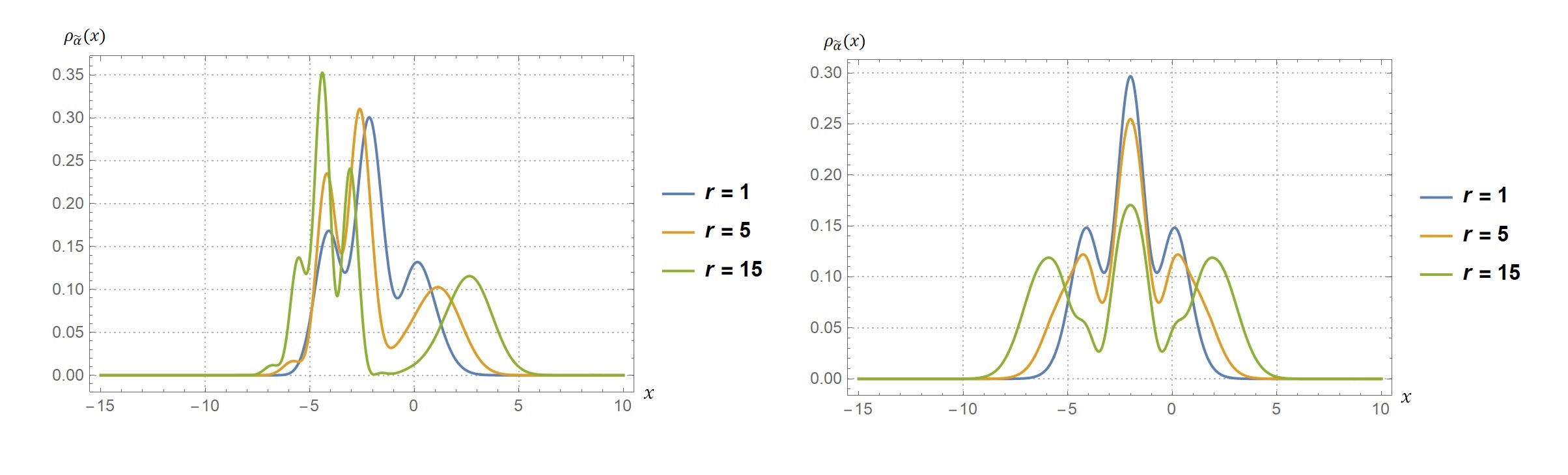}
   \caption{Probability density $\rho_{\tilde{\alpha}}(x)$ for the MCS with $f(n)=1$, $\omega_{c}^{*}=1$, $m=3$ and $j=2$. Different values of $r=\vert \tilde{\alpha} \vert$ are taken for $\theta=0$ (left) and $\theta= \pi/2$ (right). }
     \label{imagen-density32}
  \end{center}  
\end{figure}

Some graphs of $\rho_{\tilde{\alpha}}(x)$ for the MCS with $m=2$ and $m=3$ are shown in Figs. \ref{imagen-density20}-\ref{imagen-density32} for the particular case when $f(n)=1$.

%%%%%%%%%%%%%%%%%%%%%%%%%%%%%%%%%%%%%%%%%%%%%%%%%%%%%%%%%%%%%%%%%%%%%%%%%%%%%%%%%%%%%%%%%%%%%%%%%%%%%%%%%%%%%%%%%%%%%%%%%%%%%%%%%%%%%%%%%%%%%%%%%%

%%%%%%%%%%%%%%%%%%%%%%%%%%%%%%%%%%%%%%%%%%%%%%%%%%%%%%%%%%%%%%%%%%%%%%%%%
%%%%%%%%%%%%%%%%%%%%%%%%%%%%%%%%%%%%%%%%%%%%%%%%%%%%%%%%%%%%%%%%%%%%%%%%%

\subsection{Mean energy value}
In order to characterize the energy of a system, the expected value of the Hamiltonian must be calculated. For the MCS the mean energy value $\langle \hat{H}\rangle_{\tilde{\alpha}}$ is obtained through
 
\begin{equation}
\langle \hat{H}\rangle_{\tilde{\alpha}}= \langle \tilde{\alpha};m,j \vert \hat{H} \vert \tilde{\alpha};m,j \rangle,
\end{equation}
with $\hat{H}$ being the bilayer graphene Hamiltonian given by \cite{FM20}, 

\begin{equation}
\hat{H}= \hbar \omega_{c}^{*}
\left(\begin{array}{cc}
0 & \hat{b}^-{}^2 \\ \hat{b}^+{}^2 & 0\\
\end{array}
\right), \,\,\,\ \omega_{c}^{*}=\frac{e B}{m^{*} c}, \label{HAM}
\end{equation}
where $\omega_{c}^{*}$ is the cyclotron frequency for non-relativistic electrons with effective mass  $m^{*}$. Thus, it turns out that

\begin{equation}
\langle \hat{H}\rangle_{\tilde{\alpha}}=\vert C_{j}^{m}\vert^{2} \hbar \omega_{c}^{*}\left[ \sqrt{j(j-1)} + \left( \sqrt{2}\delta_{0j}+\sqrt{(\delta_{1j}+j)!} \right)^{2} \sum_{n = 1}^\infty \dfrac{[f(j)]!^{2} \, \sqrt{(mn+j)(mn+j-1)} \,\ \vert \tilde{\alpha}\vert^{2n}}{[f(mn+j)]!^{2} \, (mn+j)!}  \right]. \label{energia}
\end{equation}
\\

\begin{figure}[t]
\begin{center} 
  \includegraphics[width=16.7cm, height=5cm]{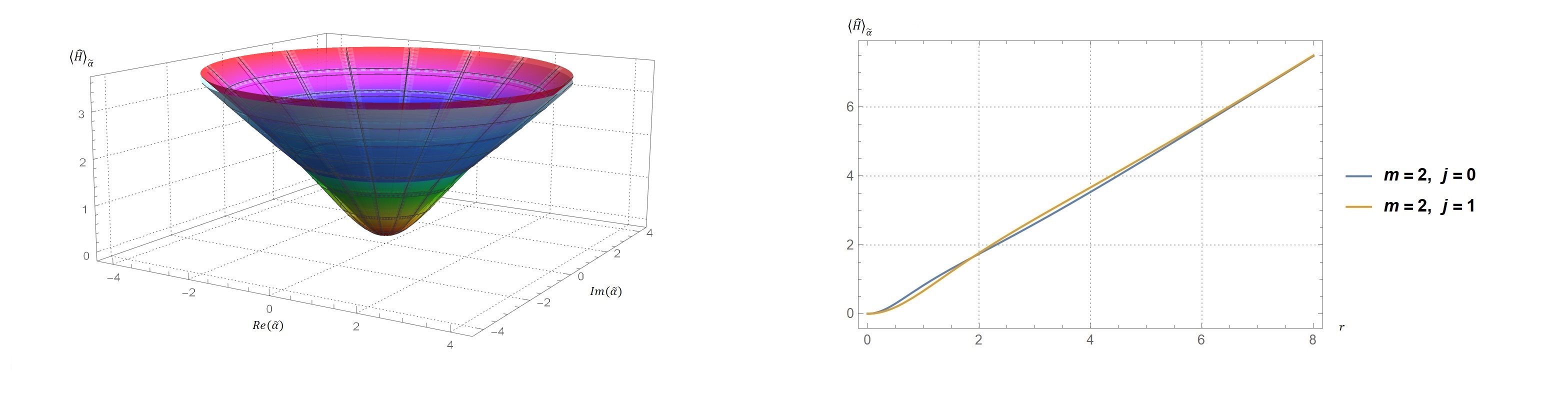}
   \caption{Mean energy value $\langle \hat{H}\rangle_{\tilde{\alpha}}$ for the MCS with $m=2$, $f(n)=1$ and $\hbar\omega_{c}^{*}=1$.}
     \label{imagen-energym2}
  \end{center}  
\end{figure}

\begin{figure}[t]
\begin{center} 
  \includegraphics[width=16cm, height=5cm]{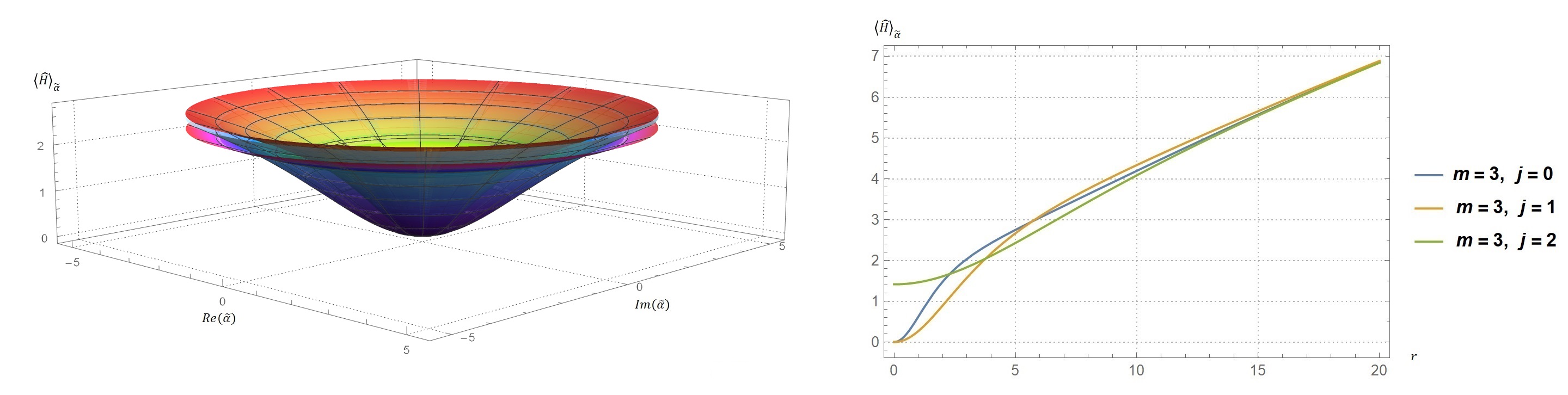}
   \caption{Mean energy value $\langle \hat{H}\rangle_{\tilde{\alpha}}$ for the MCS with $m=3$, $f(n)=1$ and $\hbar\omega_{c}^{*}=1$.}
     \label{imagen-energym3}
  \end{center}  
\end{figure}

This quantity will be useful for analyzing the time evolution of the MCS, in the same way as in \cite{FM20}. Figures \ref{imagen-energym2}-\ref{imagen-energym3} show the mean energy value (\ref{energia}) for the MCS as function of $\vert\tilde{\alpha}\vert=r$ with $f(n)=1$ and the two values of $m=\left\lbrace 2,3 \right\rbrace$.
\\

%%%%%%%%%%%%%%%%%%%%%%%%%%%%%%%%%%%%%%%%%%%%%%%%%%%%%%%%%%%%%%%%%%%%%%%%%%%%%%%%%%%%%%%%%%%%%%%%%%%%%%%%%%%%%%%%%%%%%%%%%%%%%%%%%%%%%%%%%%%%%%%%%%

%%%%%%%%%%%%%%%%%%%%%%%%%%%%%%%%%%%%%%%%%%%%%%%%%%%%%%%%%%%%%%%%%%%%%%%%%%%%%%%%%%%%%%%%%%%%%%%%%%%%%%%%%%%%%%%%%%%%%%%%%%%%%%%%%%%%%%%%%%%%%%%%%%

\subsection{Discussion}
Several physical quantities have been calculated when the system is in a MCS. As can be seen in Figs. \ref{imagen-Heisenberg2} and \ref{imagen-Heisenberg3}, the Heisenberg uncertainty relation for the MCS acquires a minimum when $\tilde{\alpha}\rightarrow0$ which depends on the eigenstate $\vert \Psi_{j}\rangle$ with the minimum energy eigenvalue involved in the corresponding expansion. Thus, for the MCS with $m=2$, $j=0$ and $m=3$, $j=0$ the HUR takes the minimum value $1/2$ while for the other three cases ($m=2$, $j=1$, $m=3$, $j=1$ and $m=3$, $j=2$) this quantity tends to $3/2$ when $\tilde{\alpha}$ goes to zero.   
\\

On the other hand, the probability density for the MCS shows an oscillatory behavior around the point $x_{0}=-2k/ \omega_{c}^{*}$, which is similar to what happens in \cite{FM20} ($\omega_{c}^{*}$ is the cyclotron frequency). Moreover, this behavior becomes more evident as $r$ increases and $\rho_{\tilde{\alpha}}(x)$ extends along the $x$-direction (see Figs. \ref{imagen-density20}-\ref{imagen-density32}). This means that, as $r$ grows, the probability to find the electron in a particular spatially confined region (in $x$-direction) decreases. Besides, when the phase of $\tilde{\alpha}$ changes the maximum value of $\rho_{\tilde{\alpha}}(x)$ also changes. Therefore, by choosing a specific $\tilde{\alpha}$ it is possible to find the electrons in a given region with the highest probability. 
\\

Finally, from Figs. \ref{imagen-energym2}-\ref{imagen-energym3} it can be seen that the mean energy value for the MCS is a continuous function of $\vert \tilde{\alpha}\vert$ whose behavior above a certain $\vert \tilde{\alpha}\vert$ is approximately linear. Moreover, when $\vert \tilde{\alpha}\vert \rightarrow 0 $ the behavior of $\langle \hat{H}\rangle_{\tilde{\alpha}}$ is different for each set of MCS, since in this limit the MCS tend to the state $\Psi_{j}$ with the minimum energy eigenvalue involved in the expansion, which is different for the MCS with different $j$ (see Eqs. (\ref{MULTI20}-\ref{MULTI32})). 
\\
\\  

\section{Evolution of the MCS for bilayer graphene}
The time evolution operator $U(t)=\mathrm{exp}(-i\hat{H}t/\hbar)$ acting on the MCS of Eq. (\ref{eq.mcs}) induces the dynamical behavior of these states, which is given by:
 
\begin{equation}
\begin{split}
\vert \tilde{\alpha};m, j; t \rangle=&C_{j}^{m}\left[ e^{-i\omega_{c}^{*}\sqrt{j(j-1)} \, t} \vert \Psi_{j} \rangle + \sum_{n = 1}^\infty \frac{\left[\sqrt{2}\delta_{0j}+\sqrt{(\delta_{1j}+j)!} \right] \, [f(j)]! \, \tilde{\alpha}^{n}}{\sqrt{(mn+j)!}\,[f(mn+j)]!} \right. \\&
\left. \times e^{-i\omega_{c}^{*}\sqrt{(mn+j)(mn+j-1)} \,t} \vert \Psi_{mn+j} \rangle\right]. \label{evol temp}
\end{split}
\end{equation}
Therefore, the evolving probability density for the MCS of bilayer graphene turns out to be  

\begin{equation}
\begin{split}
 &\rho_{\tilde{\alpha}}(x,t)=\vert C_{j}^{m}\vert^{2} \left\lbrace \dfrac{(1-\delta_{0j}-\delta_{1j})^{2}\vert \psi_{j-2}\vert^{2}+\vert \psi_{j}\vert^{2}}{\sqrt{2^{1-\delta_{0j}-\delta_{1j}}}} + \displaystyle \dfrac{{[\sqrt{2}\delta_{0j}+\sqrt{(\delta_{1j}+j)!}]}}{\sqrt{2^{2-\delta_{0j}-\delta_{1j}}}}2\Re \left[ \sum_{n=1}^\infty \frac{\tilde{\alpha}^{n}}{\sqrt{(mn+j)!}}   \right. \right. \\ &
 \left.\left. \times \frac{[f(j)]!}{[f(mn+j)]!} \left((1-\delta_{0j}-\delta_{1j})\psi_{mn+j-2} \psi_{j-2}^{*} + \psi_{mn+j}\psi_{j}^{*} \right)\exp[-i \omega_{c}^{*}(\sqrt{(mn+j)(mn+j-1)} \right. \right. \\&
 \left. \left. -\sqrt{j(j-1)}) \, t] \right] + \dfrac{[\sqrt{2}\delta_{0j}+\sqrt{(\delta_{1j}+j)!}]^{2}}{2} \left | \sum_{n=1}^\infty \frac{\exp(-i \omega_{c}^{*}\sqrt{(mn+j)(mn+j-1)} \, t) \, \tilde{\alpha}^{n}}{\sqrt{(nm+j)!}}   \right. \right. \\&
\left. \left. \times \frac{[f(j)]!}{[f(mn+j)]!} \, \psi_{mn+j-2}\right |^{2} + \dfrac{[\sqrt{2}\delta_{0j}+\sqrt{(\delta_{1j}+j)!}]^{2}}{2} \left | \sum_{n=1}^\infty \frac{\exp(-i \omega_{c}^{*}\sqrt{(mn+j)(mn+j-1)} \, t) \, \tilde{\alpha}^{n}}{\sqrt{(mn+j)!}} \right. \right. \\ & 
\left. \left. \times \frac{[f(j)]!}{[f(mn+j)]!} \, \psi_{mn+j} \right |^{2}   \right\rbrace , \label{eq.probevo1}
\end{split}
\end{equation}
which, when taking $\tilde{\alpha}=re^{i\theta}=r\left( \cos\theta+i\sin\theta\right)$ becomes

\begin{equation}
\begin{split}
 &\rho_{\tilde{\alpha}}(x,t)=\vert C_{j}^{m}\vert^{2} \left\lbrace  \dfrac{(1-\delta_{0j}-\delta_{1j})^{2}\vert \psi_{j-2}\vert^{2}+\vert \psi_{j}\vert^{2}}{\sqrt{2^{1-\delta_{0j}-\delta_{1j}}}}  + \dfrac{[\sqrt{2}\delta_{0j}+\sqrt{(\delta_{1j}+j)!}]^{2}}{2} \sum_{n=1}^\infty \sum_{k=1}^\infty \frac{r^{n+k} }{\sqrt{(mn+j)! (mk+j)!}}     \right. \\ &
 \left. \times \frac{[[f(j)]!]^{2}}{[f(mk+j)]![f(mn+j)]!} \cos[\omega_{c}^{*}(\sqrt{(mn+j)(mn+j-1)}-\sqrt{(mk+j)(mk+j-1)} \, ) \, t-(n-k)\theta] \,  \right. \\&
\left. \times (\psi_{mn+j-2} \psi_{mk+j-2}^{*}+\psi_{mn+j}\psi_{mk+j}^{*}) + \dfrac{[\sqrt{2}\delta_{0j}+\sqrt{(\delta_{1j}+j)!}]}{\sqrt{2^{2-\delta_{0j}-\delta_{1j}}}}\left[ \sum_{k=1}^\infty \frac{r^{k}}{\sqrt{(mk+j)!}}\frac{[f(j)]!}{[f(mk+j)]!}  \right. \right. \\ & 
\left. \left. \times \cos[\omega_{c}^{*}(\sqrt{(mk+j)(mk+j-1)}-\sqrt{j(j-1)} \, ) \, t-k\theta] \, ((1-\delta_{0j}-\delta_{1j})\psi_{mk+j-2}^{*} \psi_{j-2} + \psi_{mk+j}^{*} \psi_{j})  \right. \right. \\&
\left. \left. + \sum_{n=1}^\infty \frac{r^{n}}{\sqrt{(mm+j)!}}\frac{[f(j)]!}{[f(mn+j)]!} \cos[\omega_{c}^{*}(\sqrt{(mn+j)(mn+j-1)}-\sqrt{j(j-1)}) \, t-n\theta] (\left( 1-\delta_{0j}-\delta_{1j}\right)  \right. \right. \\&
\left. \left. \times \psi_{mn+j-2} \psi_{j-2}^{*} + \psi_{mn+j}\psi_{j}^{*})\right]  \right\rbrace. \label{eq.probevo11}
\end{split}
\end{equation}

\begin{figure}[!]
\begin{center} 
  \includegraphics[width=16.8cm]{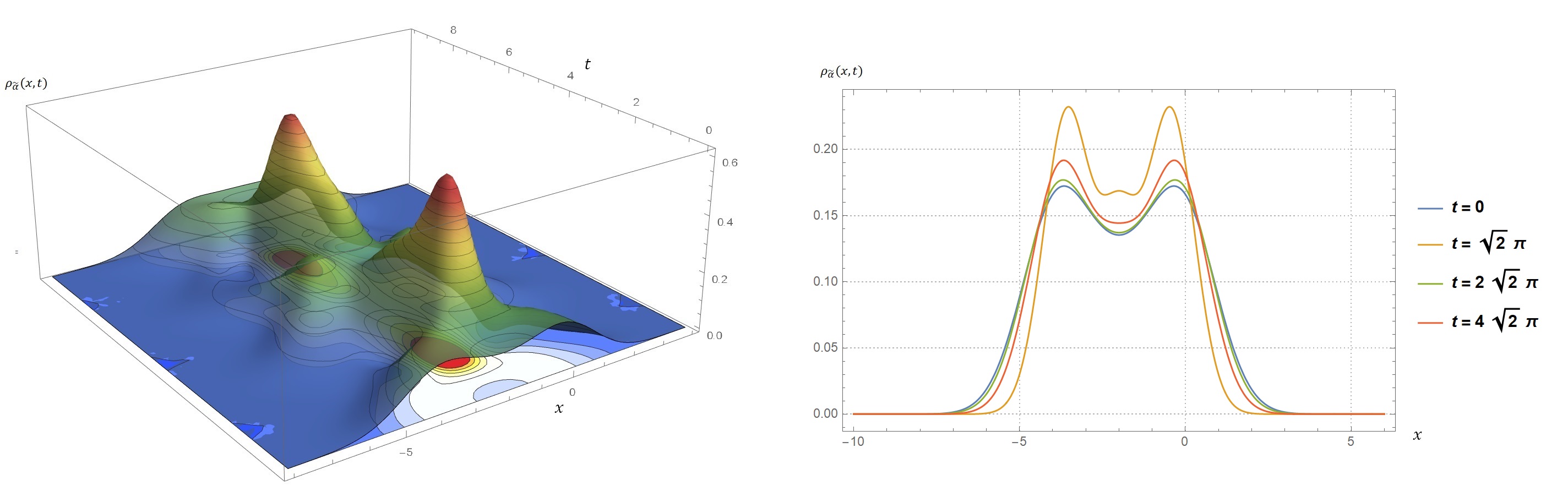}
   \caption{Left: Probability density $\rho_{\tilde{\alpha}}(x,t)$ for the bilayer graphene MCS with $f(n)=1$, $m=2$ and $j=0$. Right: Probability density $\rho_{\tilde{\alpha}}(x,t)$ at some fixed times (suggested approximate period $\tau\simeq \sqrt{2} \pi$ and some of its multiples). The values $\vert \tilde{\alpha} \vert=1$, $\theta=0$ and $\omega_{c}^{*}=1$ were taken.}
     \label{imagen-density-evol20}
  \end{center}  
\end{figure}

\begin{figure}[!]
\begin{center} 
  \includegraphics[width=16.8cm]{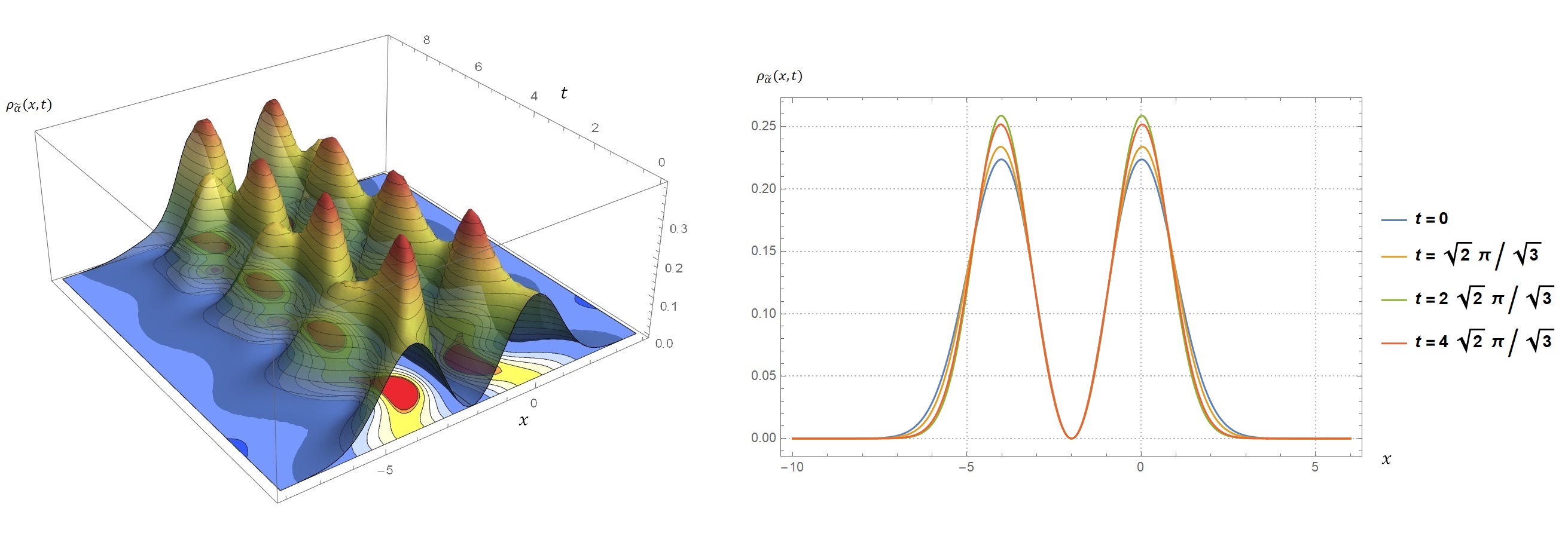}
   \caption{Left: Probability density $\rho_{\tilde{\alpha}}(x,t)$ for the bilayer graphene MCS with $f(n)=1$, $m=2$ and $j=1$. Right: Probability density $\rho_{\tilde{\alpha}}(x,t)$ at some fixed times (suggested approximate period $\tau\simeq \sqrt{2} \pi/\sqrt{3}$ and some of its multiples). The values $\vert \tilde{\alpha} \vert=1$, $\theta=0$ and $\omega_{c}^{*}=1$ were taken.}

     \label{imagen-density-evol21}
  \end{center}  
\end{figure}

\begin{figure}[!]
\begin{center} 
  \includegraphics[width=16.8cm]{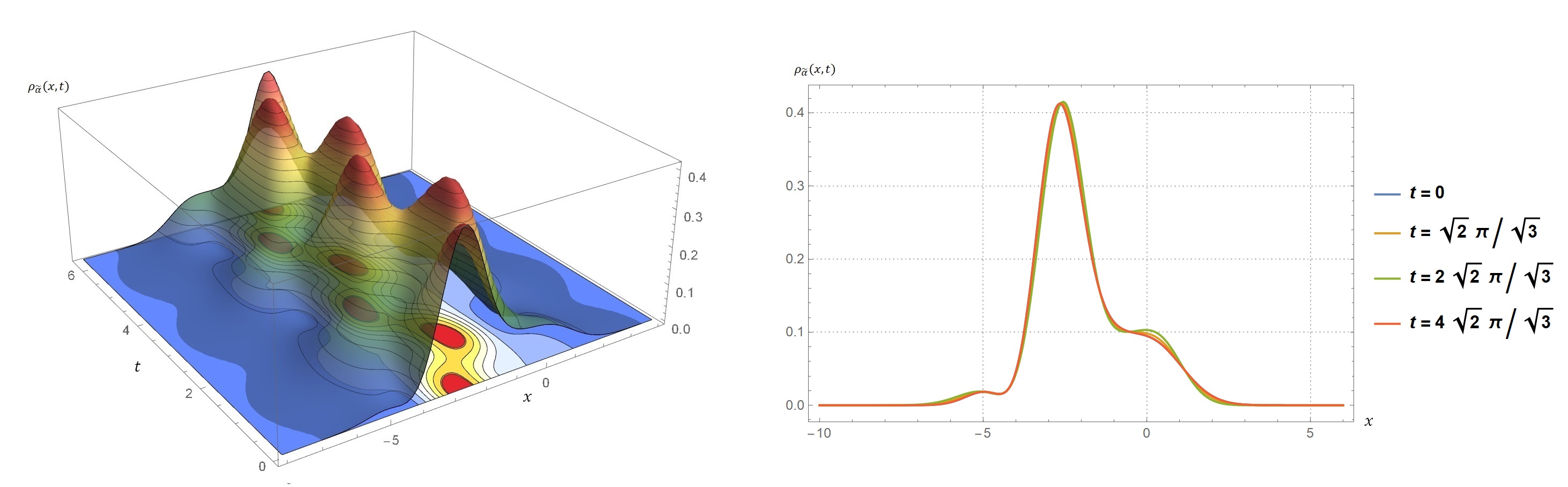}
   \caption{Left: Probability density $\rho_{\tilde{\alpha}}(x,t)$ for the bilayer graphene MCS with $f(n)=1$, $m=3$ and $j=0$. Right: Probability density $\rho_{\tilde{\alpha}}(x,t)$ at some fixed times (suggested approximate period $\tau\simeq \sqrt{2} \pi/\sqrt{3}$ and some of its multiples). The values $\vert \tilde{\alpha} \vert=1$, $\theta=0$ and $\omega_{c}^{*}=1$ were taken.}
     \label{imagen-density-evol30}
  \end{center}  
\end{figure}

\begin{figure}[!]
\begin{center} 
  \includegraphics[width=16.8cm]{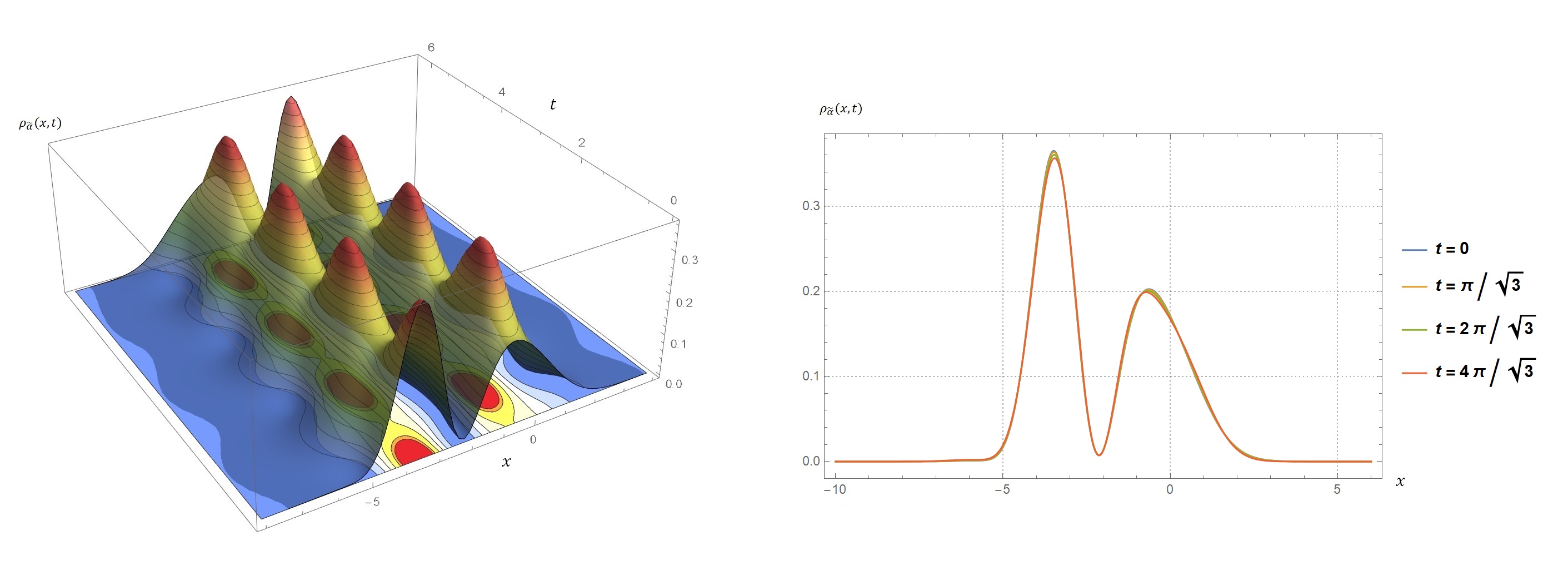}
   \caption{Left: Probability density $\rho_{\tilde{\alpha}}(x,t)$ for the bilayer graphene MCS with $f(n)=1$, $m=3$ and $j=1$. Right: Probability density $\rho_{\tilde{\alpha}}(x,t)$ at some fixed times (suggested approximate period $\tau\simeq  \pi/\sqrt{3}$ and some of its multiples). The values $\vert \tilde{\alpha} \vert=1$, $\theta=0$ and $\omega_{c}^{*}=1$ were taken.}
     \label{imagen-density-evol31}
  \end{center}  
\end{figure}

\begin{figure}[!]
\begin{center} 
  \includegraphics[width=16.8cm]{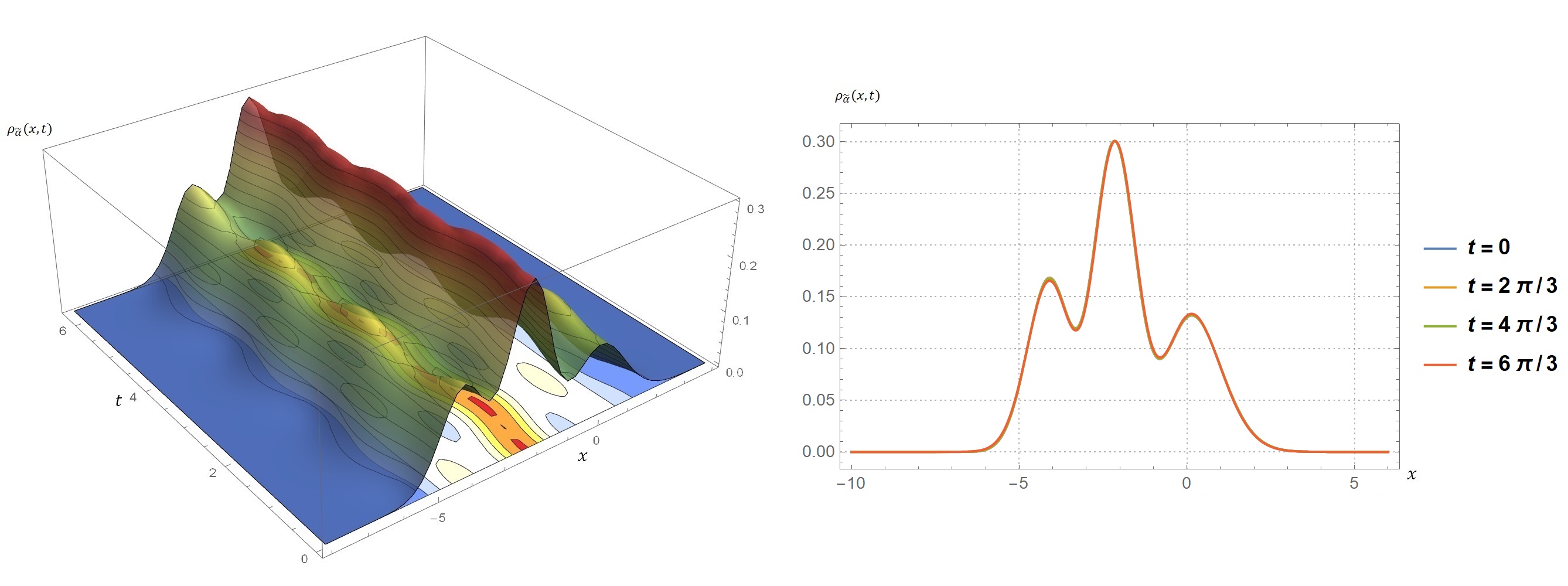}
   \caption{Left: Probability density $\rho_{\tilde{\alpha}}(x,t)$ for the bilayer graphene MCS with $f(n)=1$, $m=3$ and $j=2$. Right: Probability density $\rho_{\tilde{\alpha}}(x,t)$ at some fixed times (suggested approximate period $\tau\simeq 2 \pi/3 $ and some of its multiples). The values $\vert \tilde{\alpha} \vert=1$, $\theta=0$ and $\omega_{c}^{*}=1$ were taken.}
     \label{imagen-density-evol32}
  \end{center}  
\end{figure}

In Figs. \ref{imagen-density-evol20}-\ref{imagen-density-evol32} the probability densities for the evolving states (\ref{evol temp}) are shown, with $f(n)=1$ and the two values of $m=\left\lbrace 2,3 \right\rbrace$.
\\
\\
\\
\\

%%%%%%%%%%%%%%%%%%%%%%%%%%%%%%%%%%%%%%%%%%%%%%%%%%%%%%%%%%%%%%%%%%%%%%%%%%%%%%%%%%%%%%%%%%%%%%%%%%%%%%%%%%%%%%%%%%%%

%%%%%%%%%%%%%%%%%%%%%%%%%%%%%%%%%%%%%%%%%%%%%%%%%%%%%%%%%%%%%%%%%%%%%%%%%%%%%%%%%%%%%%%%%%%%%%%%%%%%%%%%%%%%%%%%%%%%

\subsection{Auto-correlation function}
In order to analyze further the dynamics of the MCS for bilayer graphene, the auto-correlation function $C(t)$ will be computed, as in \cite{DB20},

\begin{equation}
C(t)=\langle \Psi(t=0) \vert \Psi(t) \rangle. 
\end{equation}
Using equations (\ref{eq.mcs}) and (\ref{evol temp}) such auto-correlation function becomes

\begin{equation}
C(t)=\vert C_{j}^{m}\vert^{2} \left[ e^{-i\omega_{c}^{*}\sqrt{j(j-1)} \, t} +  \sum_{n = 1}^\infty \frac{\left[\sqrt{2}\delta_{0j}+\sqrt{(\delta_{1j}+j)!} \right]^{2} \, [[f(j)]!]^{2} \, \vert \tilde{\alpha}\vert^{2n}}{(mn+j)!\,[[f(mn+j)]!]^{2}} \, e^{-i\omega_{c}^{*}\sqrt{(mn+j)(mn+j-1)} \,t}\right]. \label{correlacion}
\end{equation}

\begin{figure}[!]
\begin{center} 
  \includegraphics[width=16.5cm]{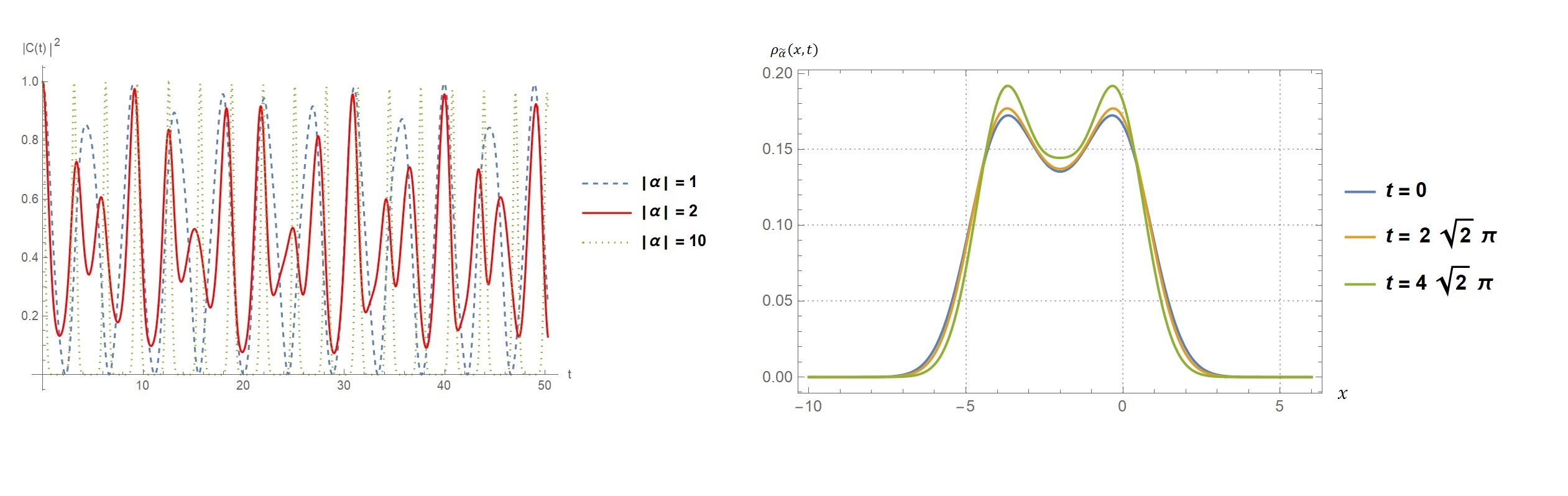}
   \caption{Left: Squared absolute value of the auto-correlation function $\vert C(t) \vert ^{2}$ for the MCS with several values of $\vert \tilde{\alpha} \vert$, $f(n)=1$, $m=2$ and $j=0$. Right: Probability density $\rho_{\tilde{\alpha}}(x,t)$ for $\vert \tilde{\alpha} \vert=1$ and several fixed times, multiples of the first approximate period $\tau_{c}\simeq 2\sqrt{2} \pi$ obtained from $\vert C(t) \vert^{2}$. The values of $\theta=0$ and $\omega_{c}^{*}=1$ were taken.}
     \label{imagen-corre20}
  \end{center}  
\end{figure}

\begin{figure}[!]
\begin{center} 
  \includegraphics[width=16.5cm]{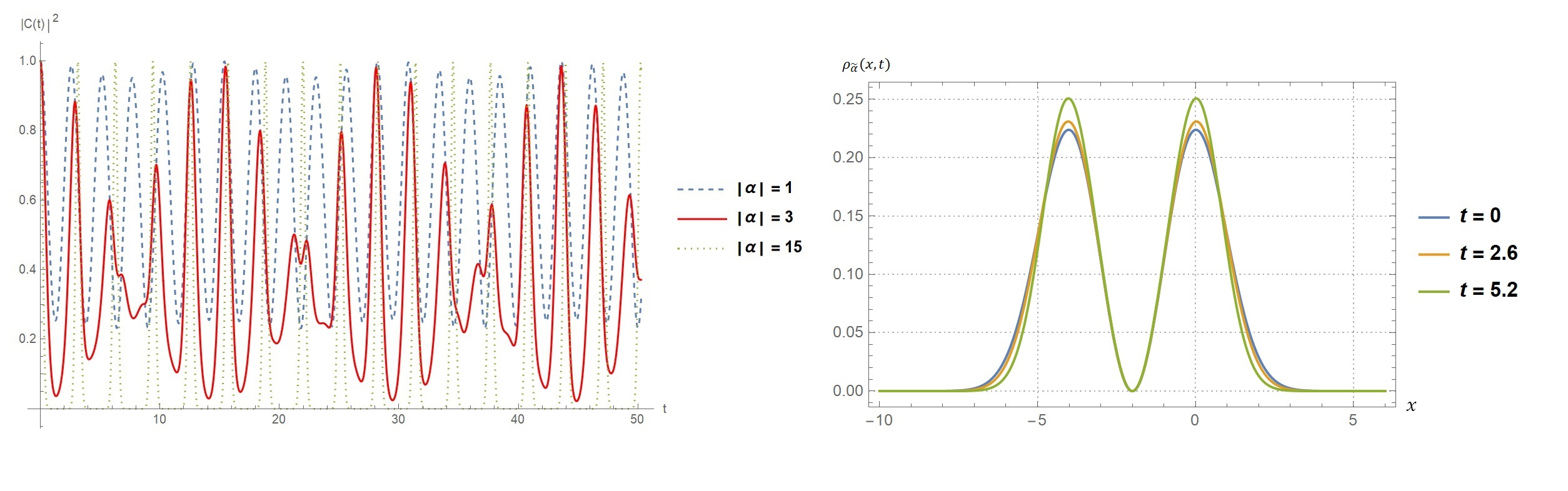}
   \caption{Left: Squared absolute value of the auto-correlation function $\vert C(t) \vert ^{2}$ for the MCS with several values of $\vert \tilde{\alpha} \vert$, $f(n)=1$, $m=2$ and $j=1$. Right: Probability density $\rho_{\tilde{\alpha}}(x,t)$ for $\vert \tilde{\alpha} \vert=1$ and several fixed times, multiples of the first approximate period $\tau_{c}\simeq 2.6$ obtained from $\vert C(t) \vert^{2}$. The values of $\theta=0$ and $\omega_{c}^{*}=1$ were taken.}
     \label{imagen-corre21}
  \end{center}  
\end{figure}

\begin{figure}[!]
\begin{center} 
  \includegraphics[width=16.5cm]{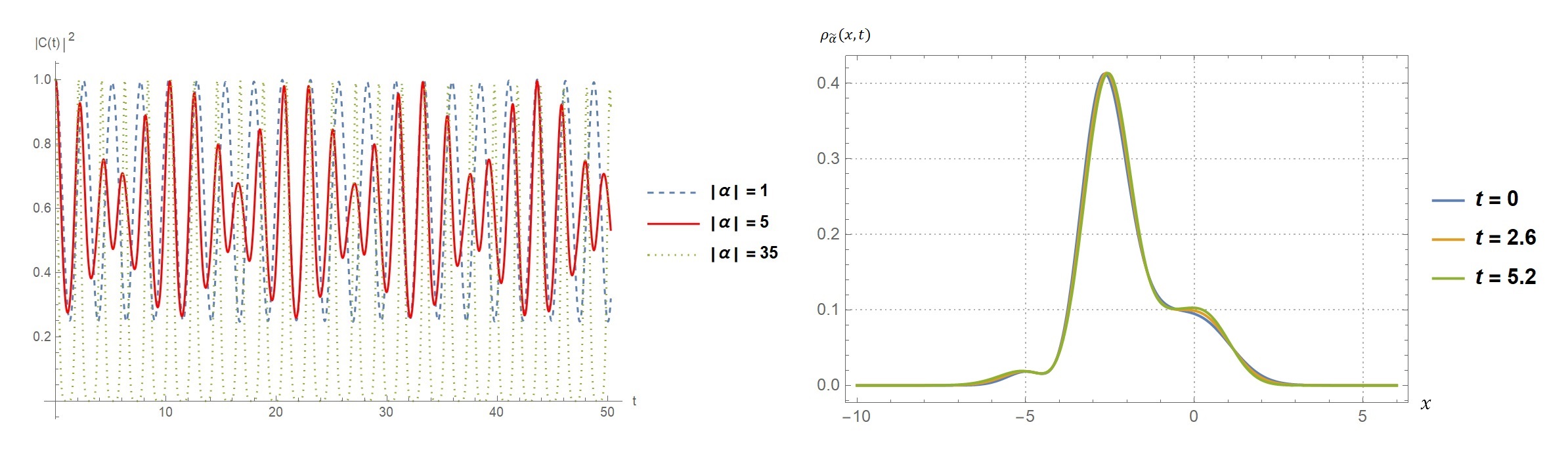}
   \caption{Left: Squared absolute value of the auto-correlation function $\vert C(t) \vert ^{2}$ for the MCS with several values of $\vert \tilde{\alpha} \vert$, $f(n)=1$, $m=3$ and $j=0$. Right: Probability density $\rho_{\tilde{\alpha}}(x,t)$ for $\vert \tilde{\alpha} \vert=1$ and several fixed times, multiples of the first approximate period $\tau_{c}\simeq 2.6 $ obtained from $\vert C(t) \vert^{2}$. The values of $\theta=0$ and $\omega_{c}^{*}=1$ were taken.}
     \label{imagen-corre30}
  \end{center}  
\end{figure}

\begin{figure}[!]
\begin{center} 
  \includegraphics[width=16.5cm]{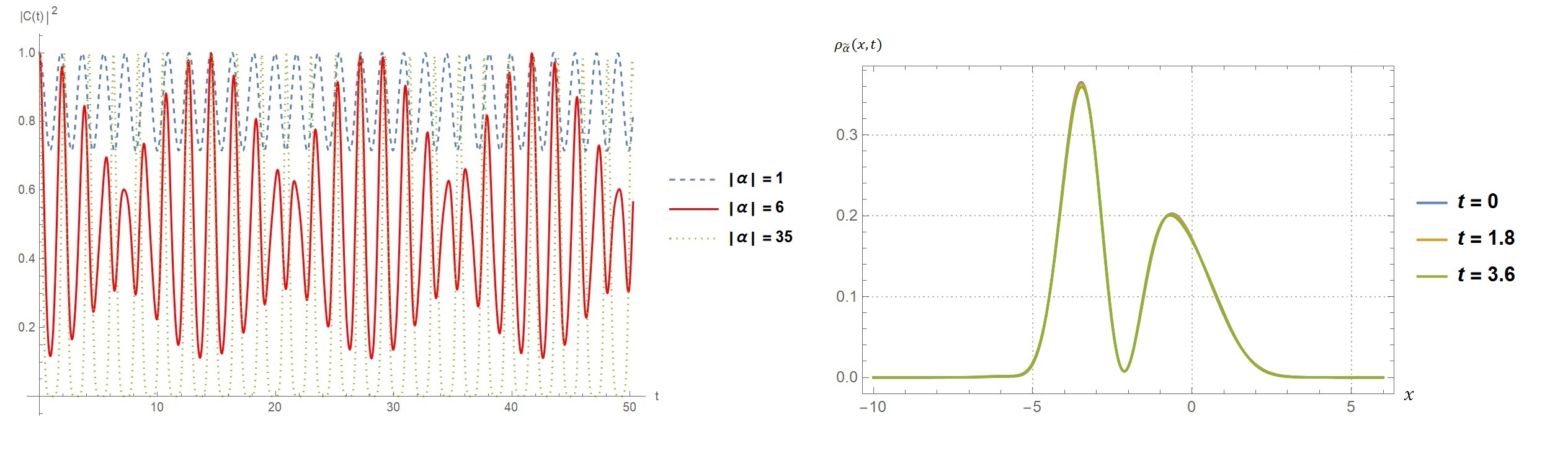}
   \caption{Left: Squared absolute value of the auto-correlation function $\vert C(t) \vert ^{2}$ for the MCS with several values of $\vert \tilde{\alpha} \vert$, $f(n)=1$, $m=3$ and $j=1$. Right: Probability density $\rho_{\tilde{\alpha}}(x,t)$ for $\vert \tilde{\alpha} \vert=1$ and several fixed times, multiples of the first approximate period $\tau_{c}\simeq 1.8 $ obtained from $\vert C(t) \vert^{2}$. The values of $\theta=0$ and $\omega_{c}^{*}=1$ were taken.}
     \label{imagen-corre31}
  \end{center}  
\end{figure}

\begin{figure}[!]
\begin{center} 
  \includegraphics[width=16.5cm]{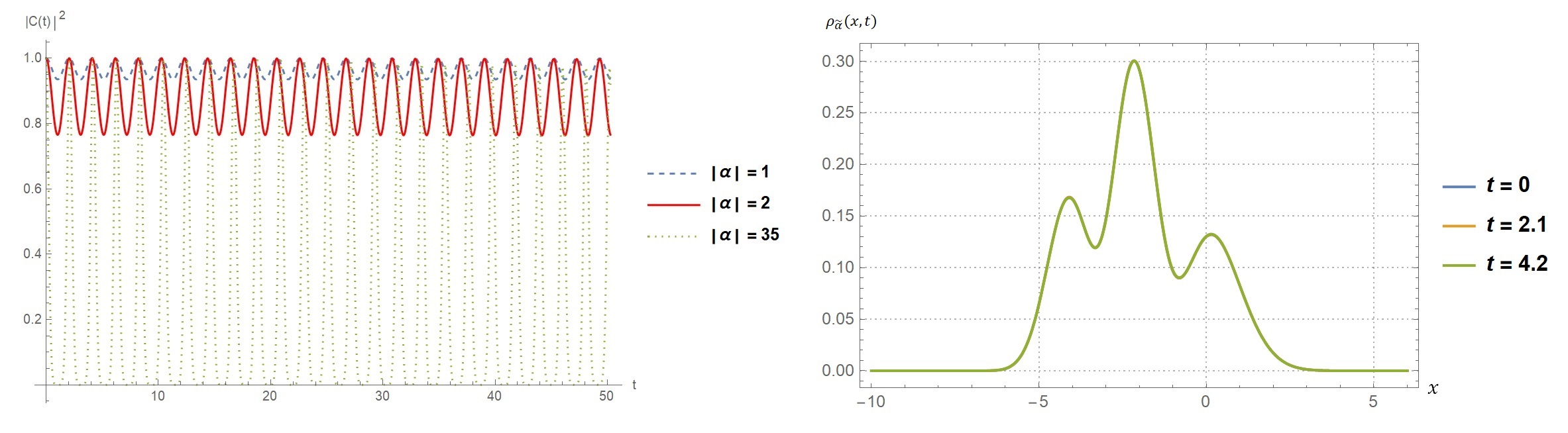}
   \caption{Left: Squared absolute value of the auto-correlation function $\vert C(t) \vert ^{2}$ for the MCS with several values of $\vert \tilde{\alpha} \vert$, $f(n)=1$, $m=3$ and $j=2$. Right: Probability density $\rho_{\tilde{\alpha}}(x,t)$ for $\vert \tilde{\alpha} \vert=1$ and several fixed times, multiples of the first approximate period $\tau_{c}\simeq 2.1$ obtained from $\vert C(t) \vert^{2}$. The values of $\theta=0$ and $\omega_{c}^{*}=1$ were taken.}
     \label{imagen-corre32}
  \end{center}  
\end{figure}

In Figs. \ref{imagen-corre20}-\ref{imagen-corre32} the squared absolute value of this auto-correlation function $\vert C(t) \vert^{2}$ for the MCS is shown, with $\omega_{c}^{*}=1$, $f(n)=1$ and several values of  $ \vert \tilde{\alpha}\vert$.

%%%%%%%%%%%%%%%%%%%%%%%%%%%%%%%%%%%%%%%%%%%%%%%%%%%%%%%%%%%%%%%%%%%%%%%%%%%%%%%%%%%%%%%%%%%%%%%%%%%%%%%%%%%%%%%%%%%%

%%%%%%%%%%%%%%%%%%%%%%%%%%%%%%%%%%%%%%%%%%%%%%%%%%%%%%%%%%%%%%%%%%%%%%%%%%%%%%%%%%%%%%%%%%%%%%%%%%%%%%%%%%%%%%%%%%%%

\subsection{Discussion}
Since the energy levels for the harmonic oscillator are equally spaced the SCS are stable in time, i.e., an SCS evolves into another SCS and for a given $\alpha$ such evolution is cyclic, with the harmonic oscillator period $\tau=2\pi / \omega$. On the other hand, for bilayer graphene the Landau-levels are not equidistant for all $n$ (see \cite{K2012,FM20}) and thus the stability in time in general cannot be guaranteed. However, as it was shown in \cite{FM20} starting from certain integer (for $n\gtrsim 2$) the energy spectrum is essentially linear (see a similar approximation in \cite{MKR94}), thus the time stability of the MCS will appear when the contribution of the eigenstates $\vert \Psi_{0} \rangle$ and $\vert \Psi_{1} \rangle$ is either small or null compared with the contribution of all other states. This behavior can be seen clearly in Fig. \ref{imagen-density-evol32}, where the evolved MCS for $m=3$ and $j=2$ are stable in time, with a period $\tau\simeq2\pi / 3\omega_{c}^{*}$. Moreover, as $\vert \tilde{\alpha}\vert$ grows ($\vert \tilde{\alpha}\vert \rightarrow \infty$) this condition is also fulfilled, thus the bilayer graphene MCS in practice are stable in time for all $m$ and $j$, with the period $\tau\simeq2\pi / m\omega_{c}^{*}$.  
\\

However, if the contribution of the states $\vert \Psi_{0} \rangle$ and $\vert \Psi_{1} \rangle$ is non-trivial compared with all other contributions the MCS $\vert \tilde{\alpha};m, j \rangle$ will be only approximately stable in time, i.e., for some values of $t$ the probability density looks similar to what it was at $t=0$ \footnote{When the evolved probability density adopts a shape similar to what it was at $t=0$ it is said that there are revivals \cite{VKTK2009}.} (see Figs. \ref{imagen-density-evol20}-\ref{imagen-density-evol31}). In order to explain the evolution of these states an approximate period $\tau$ can be calculated as previously done for the BGCS \cite{FM20}. Thus, by setting $\tilde{\alpha}$ the mean energy value is first computed, then the interval in which it lies is determined, which is bounded by two consecutive energies $E_{mn+j+m}$ and $E_{mn+j}$ such that $ E_{mn+j} < \langle \hat{H}\rangle_{\tilde{\alpha}} < E_{mn+j+m}$. Thus, the possible approximate period is obtained as follows

\begin{equation}
\tau=\dfrac{2\pi\hbar}{E_{mn+j+m}-E_{mn+j}}. \label{eq.periodo}
\end{equation}

As an example, the approximate period $\tau$ for the MCS (\ref{evol temp}) with $\vert \tilde{\alpha} \vert=1$, $m=2$ and $m=3$ has been obtained. Figs. \ref{imagen-density-evol20}-\ref{imagen-density-evol32} show the probability density $\rho_{\tilde{\alpha}}(x,t)$ for these states evaluated at this suggested period and some of its multiples.         
\\

On the other hand, as it was explained before a useful tool to analyze the dynamics of a quantum system is the auto-correlation function. This function provides a qualitative way to know how long a MCS persists at two different times. More precisely, its squared absolute value indicates how close the evolved state is to the initial state at $t=0$. For the MCS with $m=2$ and $m=3$ the auto-correlation function shows an oscillatory behavior, with an oscillation period which depends on the value of $\vert \tilde{\alpha}\vert$ (see Figs. \ref{imagen-corre20}-\ref{imagen-corre32}). If the squared absolute value of the auto-correlation function is very close or equal to one, $\vert C(t)\vert^{2} \approx 1$ \footnote{In quantum mechanics, notably in quantum information theory, a parameter called fidelity $F(\sigma, \rho)$ is defined as a measure of the distance between quantum states \cite{NC2006}. For pure states the fidelity is simply the squared absolute value of the scalar product between the two states, i.e., $F= \vert \langle \varphi_{\sigma}  \vert \varphi_{\rho}  \rangle\vert^{2}$. The reconstruction happens precisely if the fidelity is equal to one \cite{BBG2011,U2000}.}, the states $\vert \tilde{\alpha}; m, j; t \rangle$ and $\vert \tilde{\alpha}; m, j; 0 \rangle$ are said to be almost completely correlated, i.e., the MCS at $t=0$ is reconstructed for some $t>0$. Therefore, the approximate evolution period for the MCS can be determined by looking for the time $t$ when $\vert C(t)\vert^{2} \approx 1$. In Figs. \ref{imagen-corre20}-\ref{imagen-corre32} the squared absolute value of $C(t)$ is shown for several values of $\vert \tilde{\alpha} \vert$, and from these plots a suggested approximate period $\tau_{c}$ has been determined. Moreover, the probability densities for some fixed times (the suggested approximate period $\tau_{c}$ and  some of its multiples) are also plotted.   
\\

%%%%%%%%%%%%%%%%%%%%%%%%%%%%%%%%%%%%%%%%%%%%%%%%%%%%%%%%%%%%%%%%%%%%%%%%%%%%%%%%%%%%%%%%%%%%%%%%%%%%%%%%%%%%%%%%%%%%

\begin{figure}[!]
\begin{center} 
  \includegraphics[width=16.5cm]{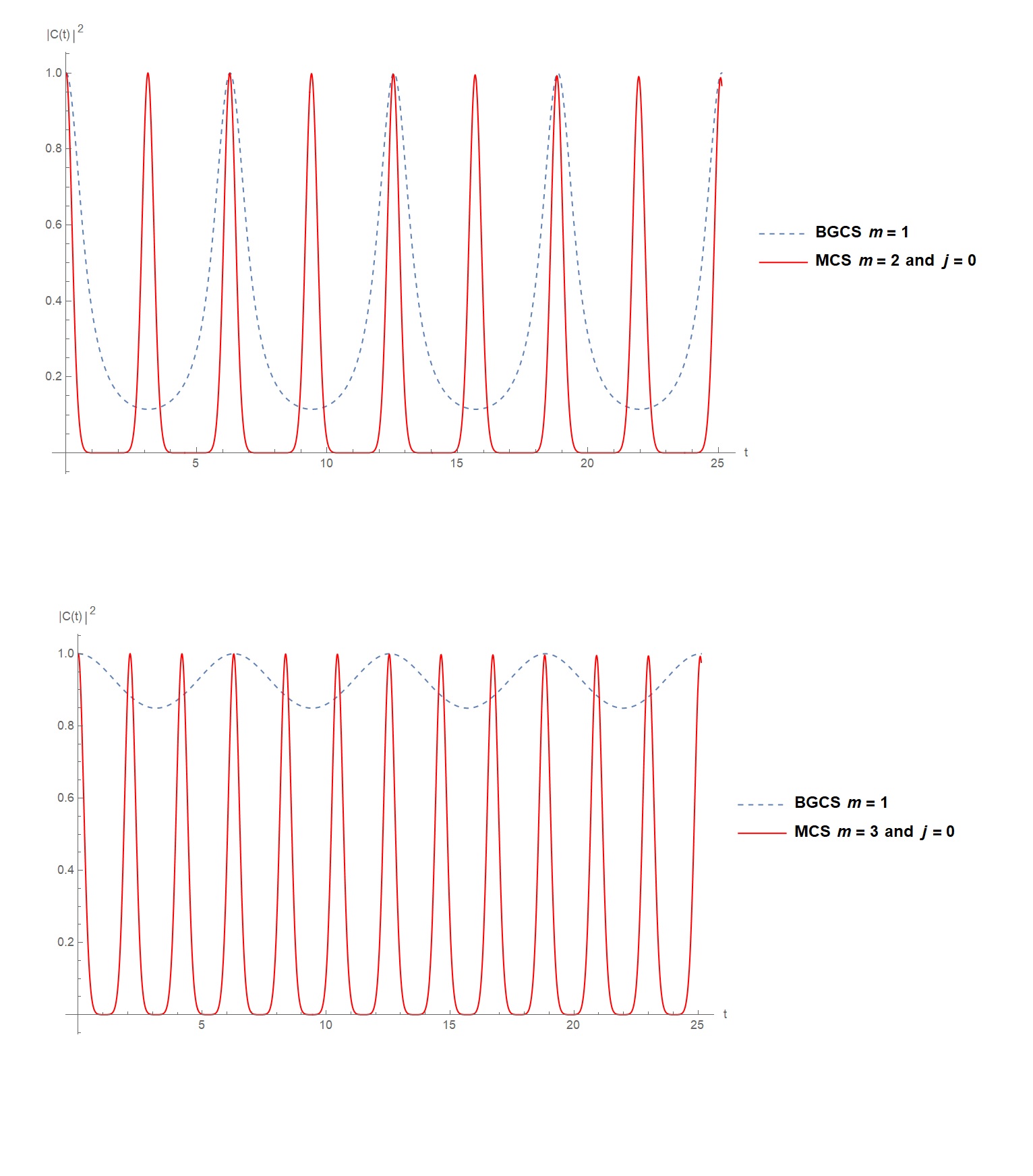}
   \caption{Up: Squared absolute value of the auto-correlation function $\vert C(t) \vert ^{2}$ for the BGCS and MCS with $f(n)=1$, $\vert \tilde{\alpha} \vert=10$, $m=2$ and $j=0$. From $\vert C(t) \vert^{2} \approx 1$ it can be seen the revivals taking place at $\tau_{BGCS}= 2\pi$ and $\tau_{MCS}= \pi$ respectively. Down: Squared absolute value of the auto-correlation function $\vert C(t) \vert ^{2}$ for the BGCS and MCS with $f(n)=1$, $\vert \tilde{\alpha}\vert=35$, $m=3$ and $j=0$. From $\vert C(t) \vert^{2} \approx 1$ it can be seen the revivals taking place at $\tau_{BGCS}= 2\pi$ and $\tau_{MCS}= 2\pi/3$ respectively. The values $\theta=0$ and $\omega_{c}^{*}=1$ were taken.}
     \label{imagen-correbgcs}
  \end{center}  
\end{figure}

%%%%%%%%%%%%%%%%%%%%%%%%%%%%%%%%%%%%%%%%%%%%%%%%%%%%%%%%%%%%%%%%%%%%%%%%%%%%%%%%%%%%%%%%%%%%%%%%%%%%%%%%%%%%%%%%%%%% 
Finally, as it was said before when the contribution of the states $\left\lbrace \Psi_{n} \right\rbrace _{n=0}^{\infty}$ with $n=0$ and $n=1$ is small compared to the eigenstates with $n\geq2$ the MCS for bilayer graphene turn out to be stable, as the BGCS derived in \cite{FM20}. Therefore, the evolved MCS and BGCS are cyclic, showing the so-called revivals in both kind of states. Moreover, in this regime the evolution period of the MCS turns out to be a fraction of the evolution period of the BGCS (see Figs. \ref{imagen-correbgcs}), i.e., $ \tau_{MCS}= \tau_{BGCS}/m=2 \pi / m \omega_{c}^{*}$, which is similar to what happens for the multiphoton coherent states of the harmonic oscillator \cite{MC19}.   

\section{Conclusions}
In this work the multiphoton coherent states were derived, in order to describe the interaction of electrons in bilayer graphene placed in a constant homogeneous magnetic field which is perpendicular to the bilayer surface. Such states are an important generalization of the CS, and constitute an alternative description allowing the quantum systems to be addressed through a semi-classical approach. Based on \cite{FM20} an appropriate generalized annihilation operator was first defined as $\hat{A}^{-}_{g}:= (\hat{A}^{-})^{m}$, then the bilayer graphene MCS were obtained as eigenstates of such operator with complex eigenvalue.
\\

In addition, in order to analyze the system some physical quantities were obtained for such states, including the Heisenberg uncertainty relation, probability density and mean energy value. It was found that in this approach the complex eigenvalue $\tilde{\alpha}$ plays an important role in the description, since it defines the system initial conditions.
\\

On the other hand, the time evolution of the MCS for bilayer graphene were studied as in \cite{FM20}. It was found that the MCS in general are not stable in time (see Figs. \ref{imagen-density-evol20} - \ref{imagen-density-evol32}), i.e., the shape of its probability density is not preserved in time since the energy spectrum of the bilayer graphene Hamiltonian $\hat{H}$ is not linear in $n$. However, since starting from certain integer ($n\gtrsim2$) the energy spectrum of $\hat{H}$ becomes practically equidistant, there are cases for which the probability density $\rho_{\tilde{\alpha}}(x,t)$ shows as well revivals suggesting that the MCS could be quasi-stable. Hence, for MCS where the contribution of the states $\vert \Psi_{0} \rangle$ and $\vert \Psi_{1} \rangle$ is small compared with the contribution of all other eigenstates, their time evolution will be quasi-stable, with a period of evolution $\tau \backsimeq 2\pi /m\omega_{c}^{*}$ being a fraction of the bilayer graphene coherent states period. Meanwhile, for the MCS where the states  $\vert \Psi_{0} \rangle$ and $\vert \Psi_{1} \rangle$ are involved in a non-trivial way, just an approximate period of evolution can be obtained through Eq. (\ref{eq.periodo}). 
\\

In this work, the auto-correlation function $C(t)$ was also derived as an additional tool to analyze the dynamical behavior of the MCS. Through its modulus squared $\vert C(t) \vert^{2}$, also called fidelity, the times $\tau_{C}$ at which the revivals of $\rho_{\tilde{\alpha}}(x,t)$ happen were obtained. Thus, despite the system does not have an equidistant energy spectrum, in this work two different ways to calculate the approximate period for the MCS have been implemented, in which the revivals of $\rho_{\tilde{\alpha}}(x,t)$ arise. 
\\

Finally, it seems possible to study similar quantum systems through this alternative approach, as the 2D Dirac materials, and to describe their dynamical behavior through the evolution of the corresponding MCS. Let us note that evolutions of this kind of systems have been recently addressed working in phase-space, by calculating the Wigner function for the corresponding coherent states \cite{BD2020,BDS2022} with the aim of looking for a way to get the first experimental realization of such functions for 2D Dirac materials.

\section{Acknowledgments}
\label{Acknowledgments}
This work was supported by CONACYT (Mexico), project FORDECYT PRONACES/61533/2020. DIMM especially acknowledge the support of CONACYT through the PhD scholarship 743766.
%%%%%%%%%%%%%%%%%%%%%%%%%%%%%%%%%%%%%%%%%%%%%%%%%%%%%%%%%%%%%%%%%%%%%%%%%%%%%%%%%%%%%%%%%%%%%%%%%%%%%%%%%%%%%%%%%%%%

\bibliographystyle{plain}

%%%%%%%%%%%%%%%%%%%%%%%%%%%%%%%%%%%%%%%%%%%%%%%%%%%%%%%%%%%%%%%%%%%%%%%%%%%%%%%%%%%%%%%%%%%%%%%%%%%%%%%%%%%%%%%%%%%%

\end{document}